\documentclass[conference]{IEEEtran}
\IEEEoverridecommandlockouts
\usepackage{optidef}
\usepackage{authblk}
\usepackage{cite}
\usepackage{booktabs}
\usepackage{tabu}
\usepackage{amsmath,amssymb,amsfonts}
\usepackage[hang,flushmargin]{footmisc}
\usepackage{multirow}
\usepackage{float}
\usepackage{textcomp}
\usepackage{lscape}
\usepackage{url}
\usepackage{enumerate}
\usepackage{caption}
\usepackage{subcaption}
\usepackage{algpseudocode}
\usepackage{graphicx}
\usepackage{textcomp}
\usepackage{xcolor}
\usepackage{array}
\usepackage{mathtools,algorithm}

\usepackage{enumitem}

\newcommand{\CASE}[1]{\STATE \textbf{case} #1\textbf{:} \begin{ALC@g}}
\newcommand{\ENDCASE}{\end{ALC@g}}

\newcommand{\DEFAULT}{\STATE \textbf{default:} \begin{ALC@g}}
\newcommand{\ENDDEFAULT}{\end{ALC@g}}
\newcommand{\DEFAULTLINE}[1]{\STATE \textbf{default:} }
\newcolumntype{L}[1]{>{\raggedright\let\newline\\\arraybackslash\hspace{0pt}}m{#1}}
\newcolumntype{C}[1]{>{\centering\let\newline\\\arraybackslash\hspace{0pt}}m{#1}}
\newcolumntype{R}[1]{>{\raggedleft\let\newline\\\arraybackslash\hspace{0pt}}m{#1}}
\def\BibTeX{{\rm B\kern-.05em{\sc i\kern-.025em b}\kern-.08em
    T\kern-.1667em\lower.7ex\hbox{E}\kern-.125emX}}

\begin{document}

\title{Optimal Slot Size under Various Bandwidth Distributions in the Flexible-grid Optical Networks}

\author[ ]{Varsha Lohani}
\author[ ]{Anjali Sharma}
\author[ ]{Yatindra Nath Singh} 

\affil[  ]{Department of Electrical Engineering, Indian Institute of Technology Kanpur, Kanpur, India}

\affil[  ]{\textit {lohani.varsha7@gmail.com*, anjalienix05@gmail.com and ynsingh@iitk.ac.in}}

\maketitle
\begin{abstract}
Flexible grid Optical Networks are efficient mechanism to provide flexibility in the optical spectrum utilization. For such networks, the slot width size as specified by the ITU-T G.694.1 is 12.5 GHz. However, one should question if it is the optimal grid size? In this paper, under different bandwidth distribution scenarios, we review which slot size give appropriate spectrum efficiency. Moreover, we present a study of the slot sizes with varying incoming traffic having some bandwidth requirement under different scenarios.
\end{abstract}

\section{Introduction}

Due to the proliferation of various services, including multimedia and cloud computing applications, there is enormous growth in Internet traffic in the last many years. According to the Cisco Annual Internet Survey, nearly two-thirds of the world's population will have Internet connectivity by 2023. The number of nodes connected to the IP networks would be more than three times the number of people worldwide.  In 2023, the average of fixed global broadband speeds will hit 110.4 Mbps, up from 45.9 Mbps in 2018\cite{cisco}. The evolution of high-speed technology requires a network that can satisfy the bandwidth requirements. In the past few decades, two such networks built upon optical fibre technologies have been proposed and studied. These are the Fixed-grid optical networks and Flexible-grid optical networks.

The electromagnetic spectrum range from 850 nm through 1550 nm is commonly available for use in glass fibres. Opting the entire optical fibre bandwidth for setting up a single lightpath in optical networks results in inefficient spectrum usage. The interface electronics also limit the single lightpath capacity, which is much less than the available bandwidth. Therefore, the bandwidth in optical networks needs to be harnessed using various Multiplexing Techniques. In Wavelength Division Multiplexing (WDM) based Optical Networks (Fixed-grid), and Elastic Optical Networks (Flexible-grid) the whole optical bandwidth is divided into multiple slots (parallel sub-channels) of different granularities. These slots are used for multiple data transmissions parallelly.

\begin{figure}[h]
    \centering
    \includegraphics[width=\linewidth]{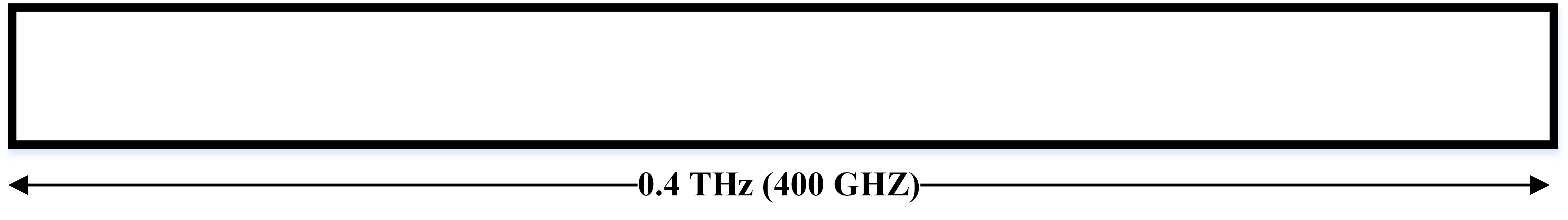}
    \caption{An optical fiber link of bandwidth 400 GHz.}
    \label{fig:bw}
\end{figure}

\begin{figure}[h]
    \centering
    \includegraphics[width=\linewidth]{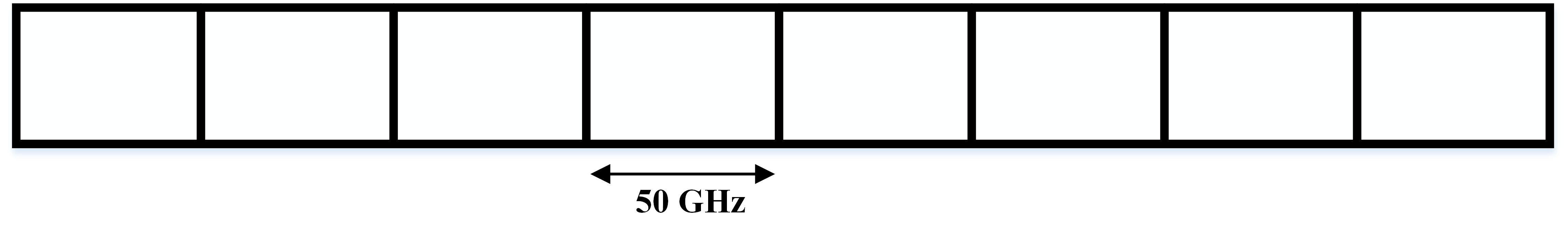}
    \caption{Partitioning of optical fiber link of bandwidth 400 GHz (figure \ref{fig:bw}) using eq. \ref{eqn0} into 8 slots where the slot width is 50 GHz.}
    \label{fig:wdm}
\end{figure}

\begin{figure}[h]
    \centering
    \includegraphics[width=\linewidth]{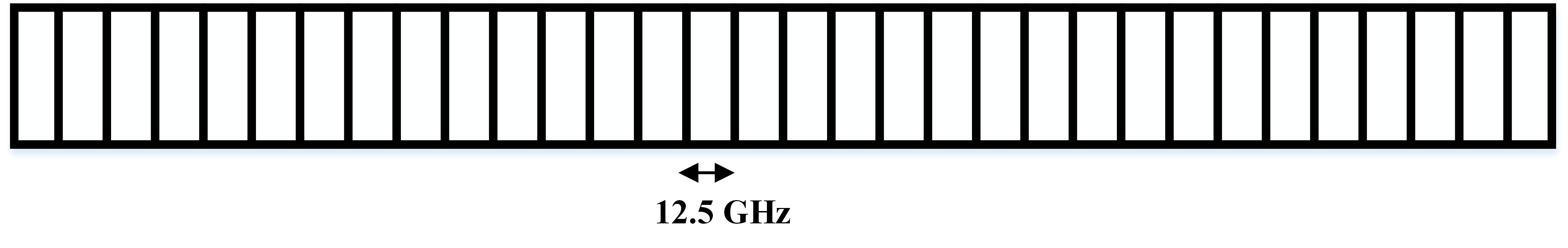}
    \caption{Partitioning of optical fiber link of bandwidth 400 GHz (figure \ref{fig:bw}) using eq. \ref{eqn0} into 32 slots where the slot width is 12.5 GHz.}
    \label{fig:ofdm}
\end{figure}

With the help of an optical network representation, we will explain the slot width problem. We can represent an optical network as a graph \textit{G(N, L)}, where \textit{G} consists of a set of optical nodes \textit{N}, indexed by \textit{n} and a set of optical fiber links \textit{L}, indexed by \textit{l}. Each optical link, \textit{l}, is connected to a pair of vertices (\textit{$l(i, j) \in L$}, where $i$ and $j \in N$) and has usable bandwidth, $B_{l}$. In order to use links efficiently, the $B_{l}$ is segmented into multiple spectrum slots. The cardinality of the possible spectrum slots on an link \textit{l}, $|\Delta_{l}|$  is represented by
    \begin{equation}
    |\Delta_{l}| = \Bigl \lfloor{\dfrac{\text{$B_l$ (GHz)}}{\text{Slot Width (GHz)}}}\Bigr \rfloor.
    \label{eqn0}
    \end{equation}
The optical spectrum in the fixed grid network is divided into 50 GHz or 100 GHz equal-width slots as shown in figure \ref{fig:wdm}. Sometimes, there is no room for connection requests with bandwidth higher than channel-width, even after using suitable modulation formats. Also, connection requests with low bandwidth requirement lead to wastage as the whole wavelength channel needs to be allocated irrespective of the requested bandwidth. It decreases the overall spectrum efficiency of the links and the network. 
Hence, to improve the utilization of spectrum, the slots allocated to the requests were made flexible. In a flexible grid optical networks, the slot width is reduced from 50 GHz to 12.5 GHz \cite{itut} as shown in figure \ref{fig:ofdm}. This reduction provides support for higher bandwidth demands \cite{eon}. However, for multiple slot allocation, the contiguity and continuity constraints in the network need to be followed. Dynamically arriving and departing connections results in fragmentation within the network spectrum. This leads to low network utilization and a higher number of blocked connections. 
To overcome the inefficiencies in the WDM-Optical Networks, the Elastic Optical networks, which is also called Flexible-Grid network, has been the focus of study over the last decade. In most of the research works, the slot width for Flexible-grid optical networks was assumed to be 12.5 GHz. In \cite{sw1} and \cite{sw2}, the slot width performance analysis is done for multiples of 12.5 GHz. The bandwidth distribution in these papers are assumed to be uniform\footnote{The probability of each incoming demand rate is same.}. A natural question arises that whether 12.5 GHz is the optimal slot size for dynamic traffic scenarios? If not, then how to determine the appropriate slot size? 

One solution is to take different bandwidth distributions and calculate the spectrum efficiency\textit{(b/s/Hz)} for various slot sizes. In this paper, for various performance parameters, i.e., Blocking Probability, Bandwidth Blocking Probability and spectrum Efficiency, we determine the slot width that is most suitable for various bandwidth distributions. 


\section{Performance Analysis of Different Slot Widths for Various Bandwidth Distributions}

The slot size study done in this paper also considers the type of networks along with the Routing and Spectrum Assignment (RSA) algorithms used. Analysis is done based on the distribution of the incoming demands such that maximum utilization of the spectrum can be achieved. 

\subsection{Network Settings}

\begin{figure}
\begin{subfigure}{0.4\textwidth}
    \centering
    \includegraphics[width=\linewidth]{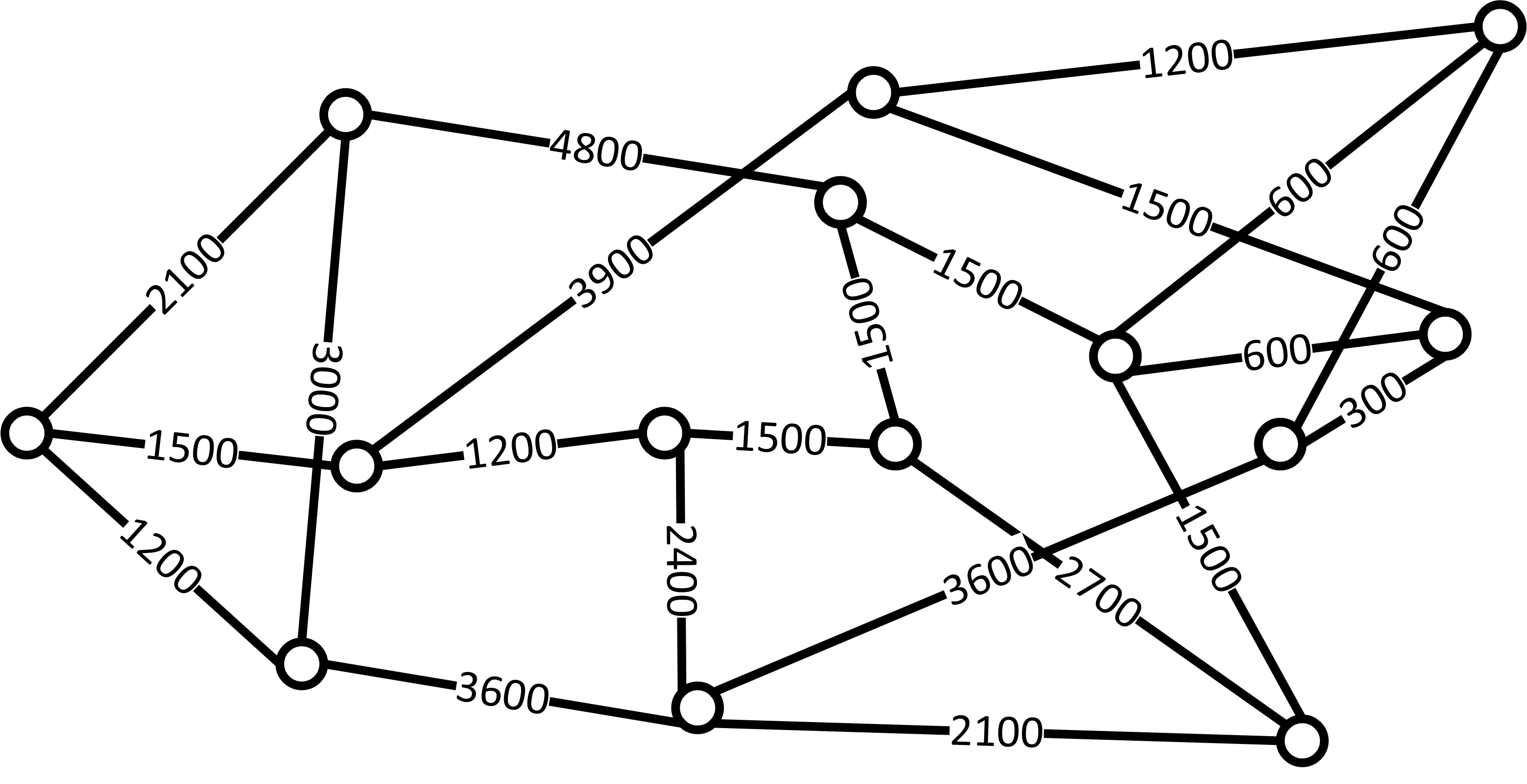}
    \caption{14 nodes, 22 links NSFNET.}
    \label{fig:nsfnet}
\end{subfigure}
\begin{subfigure}{0.4\textwidth}
    \centering
    \includegraphics[width=\linewidth]{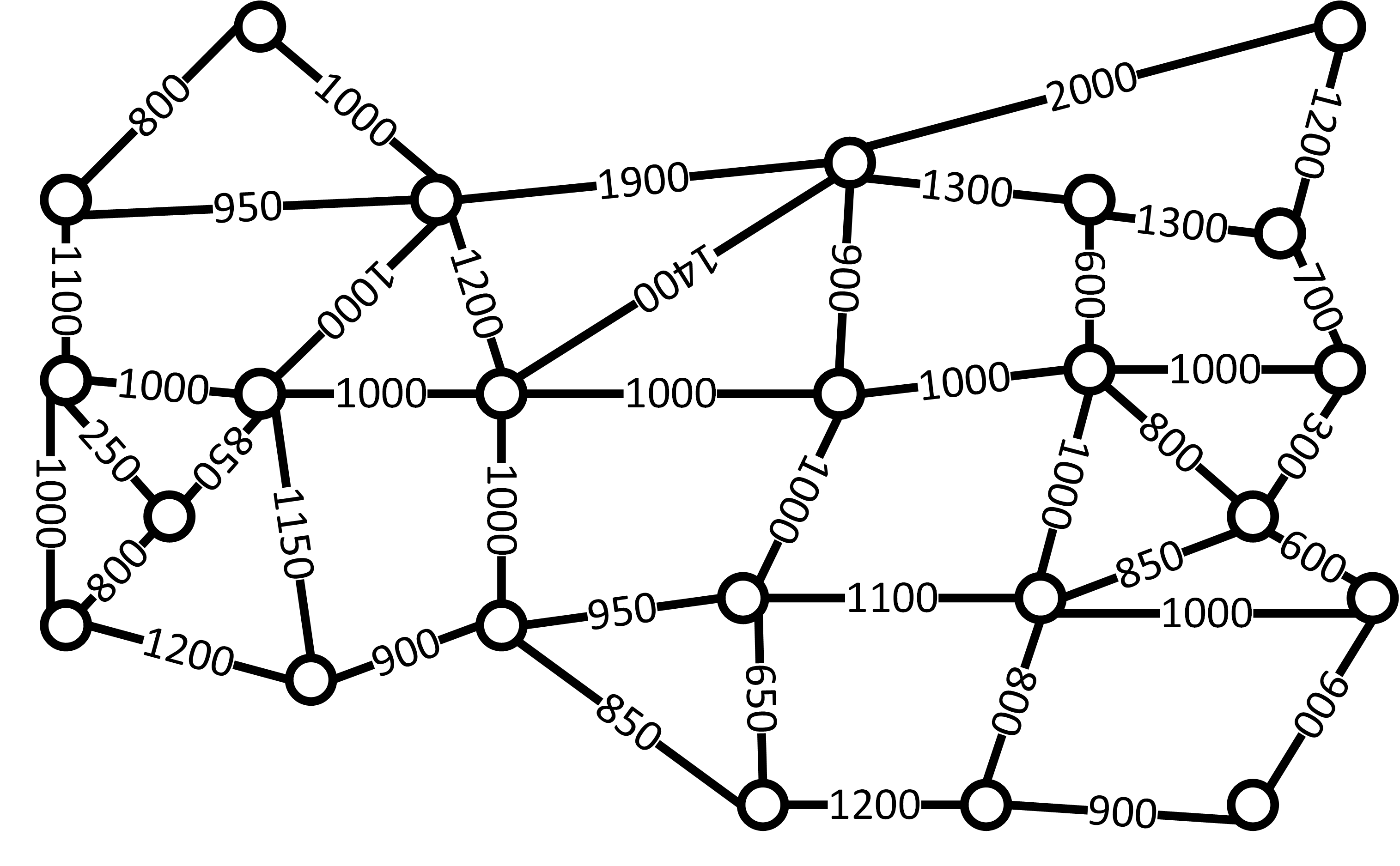}
    \caption{24 nodes, 43 links USNET.}
    \label{fig:usnet}
\end{subfigure}
\begin{subfigure}{0.4\textwidth}
    \centering
    \includegraphics[width=0.3\linewidth]{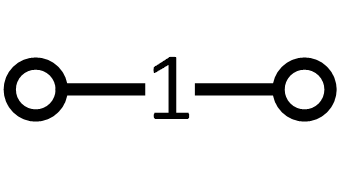}
    \caption{Single link.}
    \label{fig:sl}
\end{subfigure}
\caption{Networks Topology used, with distance in km marked on the each edge.}
\label{fig:nt2}
\end{figure}

We operated a set of simulation experiments using MATLAB R2019b. The simulations are done for 2,00,000 requests over multiple iterations. The performance of the proposed method is evaluated on the 14-nodes 22-links NSFNET with an average nodal degree\footnote{a.k.a. nodal degree, $n_d = \dfrac{2E}{V}$} of 3.0, 24-nodes 43-links USNET with an average nodal degree of 3.5, and 2-nodes single link network as shown in figure \ref{fig:nt2}. We assume the fiber bandwidth to be 4 THz on each link of the network. We have taken slot widths from 2 GHz onwards upto 100 GHz (2 $\leq$ \textit{n} $\leq$ 100). Using Optical-Orthogonal Frequency Division Multiplexing (O-OFDM) technology, the whole bandwidth is divided into \textit{n} GHz parallel channels. The bandwidth required by each lightpath request follows different bandwidth distributions depending on bandwidth parameter characterizing it. In this paper, we consider three types of bandwidth distributions for generation of required bandwidth required by various lightpath requests.

\begin{enumerate}

\item \textbf{Uniform Distribution of Bandwidth:} The bandwidth $B$ (in Gbps) required by each lightpath requests is distributed uniformly from $B_{min}$ to $B_{max}$, i.e., with equal probability, bandwidth of each incoming lightpath can take any value in between $B_{min}$ and $B_{max}$. 

\item \textbf{Poisson Distribution of Bandwidth:} The bandwidth required by each lightpath requests is characterized by average bandwidth ($B_{avg}$ Gbps). Let, $\Delta f$ be the granularity, independent of the slot size and should be as small as possible. We have assumed $\Delta f$ to be 1 MHz. Therefore,  

\begin{equation*}
   B'_{avg}= \dfrac{B_{avg}}{\Delta f},
\end{equation*}

where $B'_{avg}$ is the average bandwidth in terms of number of bandwidth granules.

We define Poisson distribution with parameter $B'_{avg} > 0$. Here, we considered a discrete random variable ${Y = k}$, where $k$ is the required bandwidth (in terms of number of bandwidth granules) by a lightpath request at a specific interval and it follows the Poisson distribution. Also, $k$ can be either $>$ or $<$ or $=$ $B'_{avg}$. The Probability Mass Function (PMF) for Poisson distribution is: 

\begin{equation*}
f(k; B'_{avg}) = Pr(Y = k) = \dfrac{(B'_{avg})^{k} \times e^{-B'_{avg}}}{k!}.
\end{equation*}

We are using MATLAB's inbuilt function to generate the values of \textit{k}. Then, corresponding $B_{req}$ (in Gbps) can be computed using,

\begin{equation*}
   B_{req} = k \times {\Delta f}.
\end{equation*}

\item \textbf{Constant Bandwidth:} The bandwidth required by each lightpath requests is $B$ Gbps. 
\end{enumerate}

For spectrum allocation, we also considered an additional guard band (GB). The size of GB is considered to be 10 GHz. 

The incoming lightpath requests arrive with exponentially distributed inter-arrival times with the average time between consecutive arrivals as $\dfrac{1}{\lambda}$ seconds and each connection is maintained for an exponentially distributed holding time with average of $\dfrac{1}{\mu}$ seconds before being released. The offered load per node ($\rho$) in Erlang (E) is given by

\begin{equation*}
\rho = \dfrac{\dfrac{1}{\mu}}{\dfrac{1}{\lambda}} = \dfrac{\lambda}{\mu} \quad \textrm{Erlangs/node}
\end{equation*} 
The performance parameters are estimated based on the observations made during the steady-state condition (approximately after three times the average holding which observed to be happening, i.e., $3 \times \dfrac{1}{\mu}$).


Intuitively, the smallest slot size\footnote{Single slot.} should give the best performance, e.g., if the traffic range is from 1 Gbps to $B_{max}$ Gbps\footnote{$B_{max}>1$}, where $B_{max}$ is the maximum allowable traffic, then the slot size of 1 GHz provides $100\%$ spectrum efficiency. Nevertheless, a narrow slot size requires additional overhead in the network and also leads to more fragmentation induced blocking.

\subsection{Performance Metrics} 
The performance of the proposed method is evaluated on the basis of blocking probability of the incoming requests, blocking probability of the incoming required slots, and the spectrum efficiency, with the gradual increments in the demand rate (in Gbps): 
    \begin{itemize}
        \item \textbf{Blocking Probability}: It is defined as the ratio of total number of blocked connections to the total number of arrived connections.
        \item \textbf{Bandwidth Blocking Probability}: It is defined as the ratio of the total amount of incoming bandwidth or slots requests which are blocked, to the total amount of bandwidth or slots required by all the incoming connections.
        \item \textbf{spectrum Efficiency}:  spectrum Efficiency is defined as the ratio of the actual bandwidth used by the requests irrespective of allocated slots to the total Bandwidth allocated in the network. The total bandwidth allocated by way of assigned spectrum slots is not fully used as shown in fig. \ref{fig:se}.

        \begin{figure}[h]
            \centering
            \includegraphics[width=0.8\linewidth]{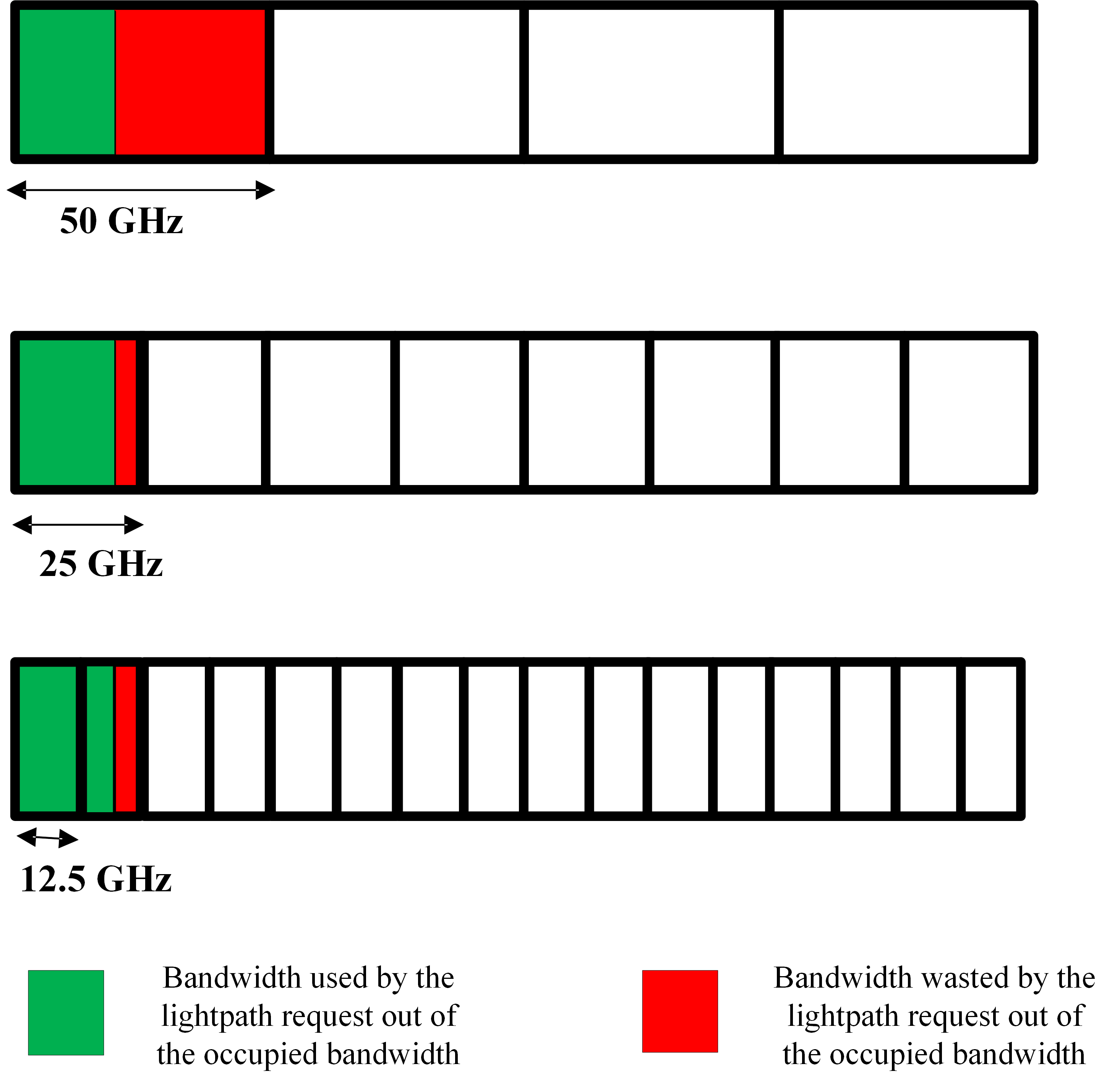}
            \caption{A connection request of 50 Gbps is allocated in Flexible-grid link of different slot sizes.}
            \label{fig:se}
        \end{figure}

        The spectrum slots or sub-bandwidth can be either occupied or unoccupied by the lightpath requests. The occupied ones are either completely used or partly used. If they are completely used, then there is efficient use of spectrum. If they are partly used, then there is wastage of bandwidth resulting in spectrum inefficiency.

\end{itemize}
\subsection{Illustrative Numerical Analysis}

In this section, for different bandwidth distributions, we are going to evaluate the various performance metrics for a range of slot widths. This will allow us to determine appropriate slot width to be used in Flexible-grid Optical Networks.

\begin{figure}[h]
     \centering
     \captionsetup{justification=centering}

     \begin{subfigure}[b]{0.2\textwidth}
         \centering
         \includegraphics[width=\linewidth]{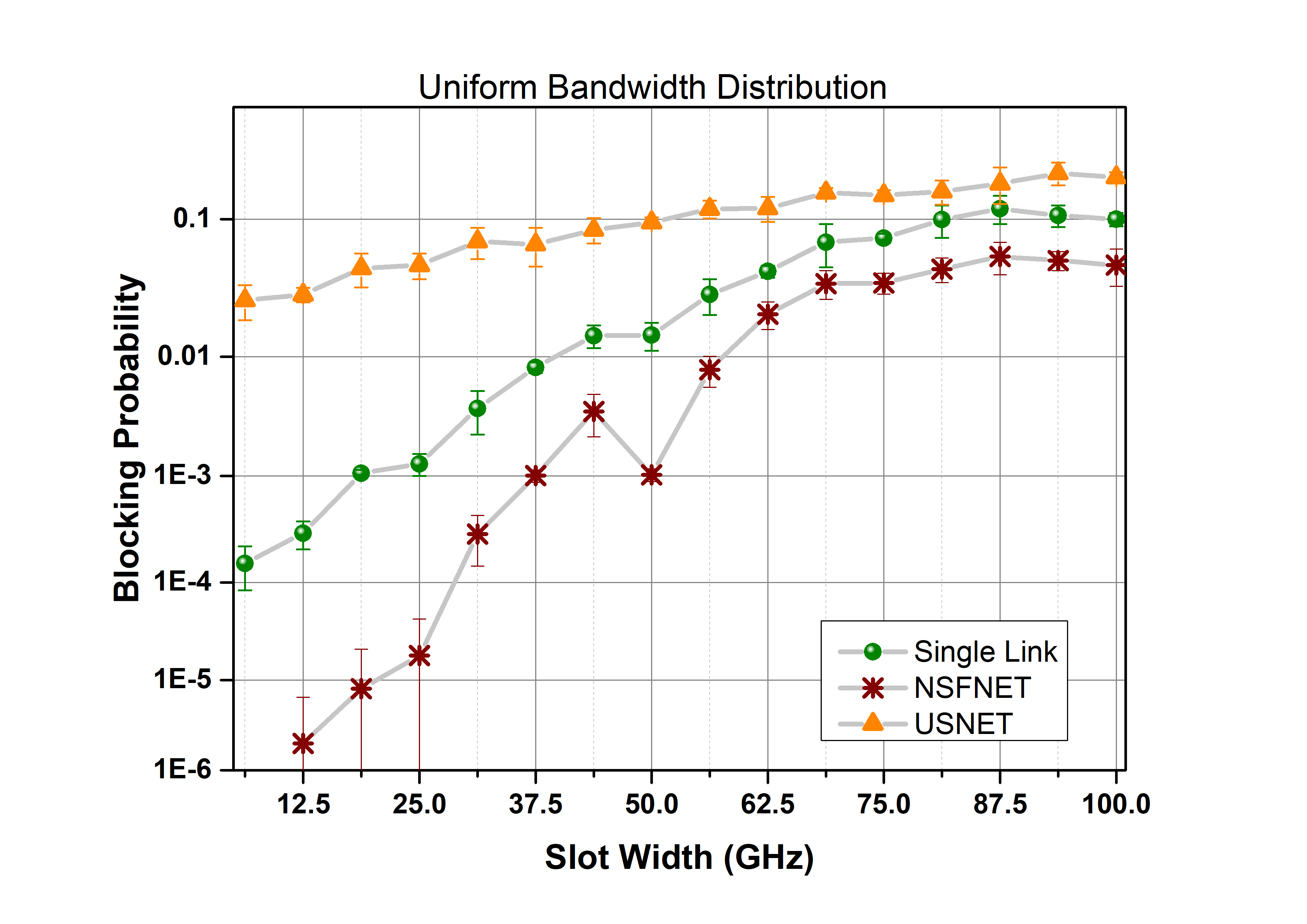}
         \caption{Blocking Probability against various Slot Width.}
         \label{fig:bpun}
     \end{subfigure}
     \hfill
     \begin{subfigure}[b]{0.2\textwidth}
         \centering
         \includegraphics[width=\linewidth]{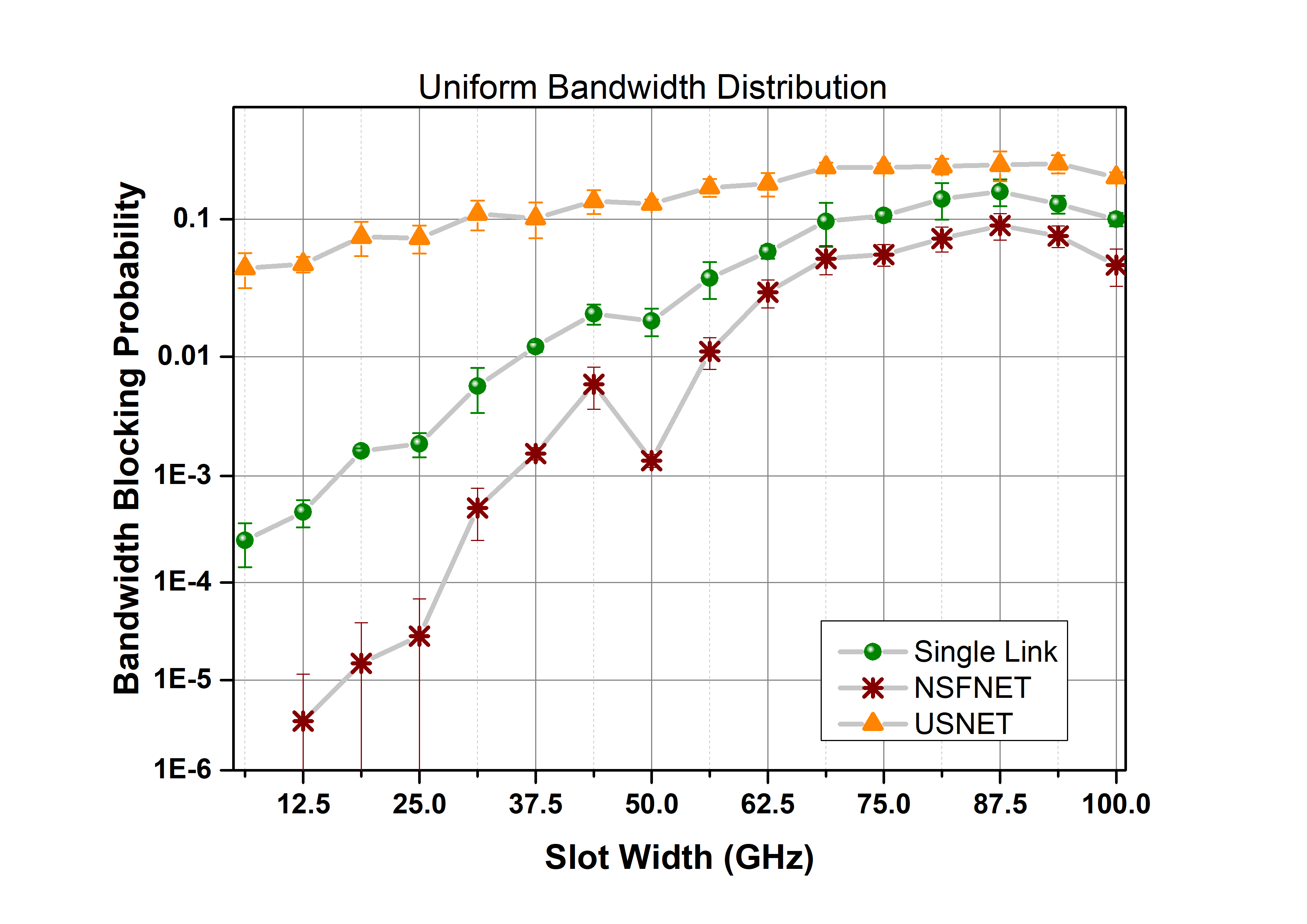}
         \caption{Bandwidth Blocking Probability against various Slot Width.}
         \label{fig:bbpun}
     \end{subfigure}
     \begin{subfigure}[b]{0.2\textwidth}
         \centering
         \includegraphics[width=\linewidth]{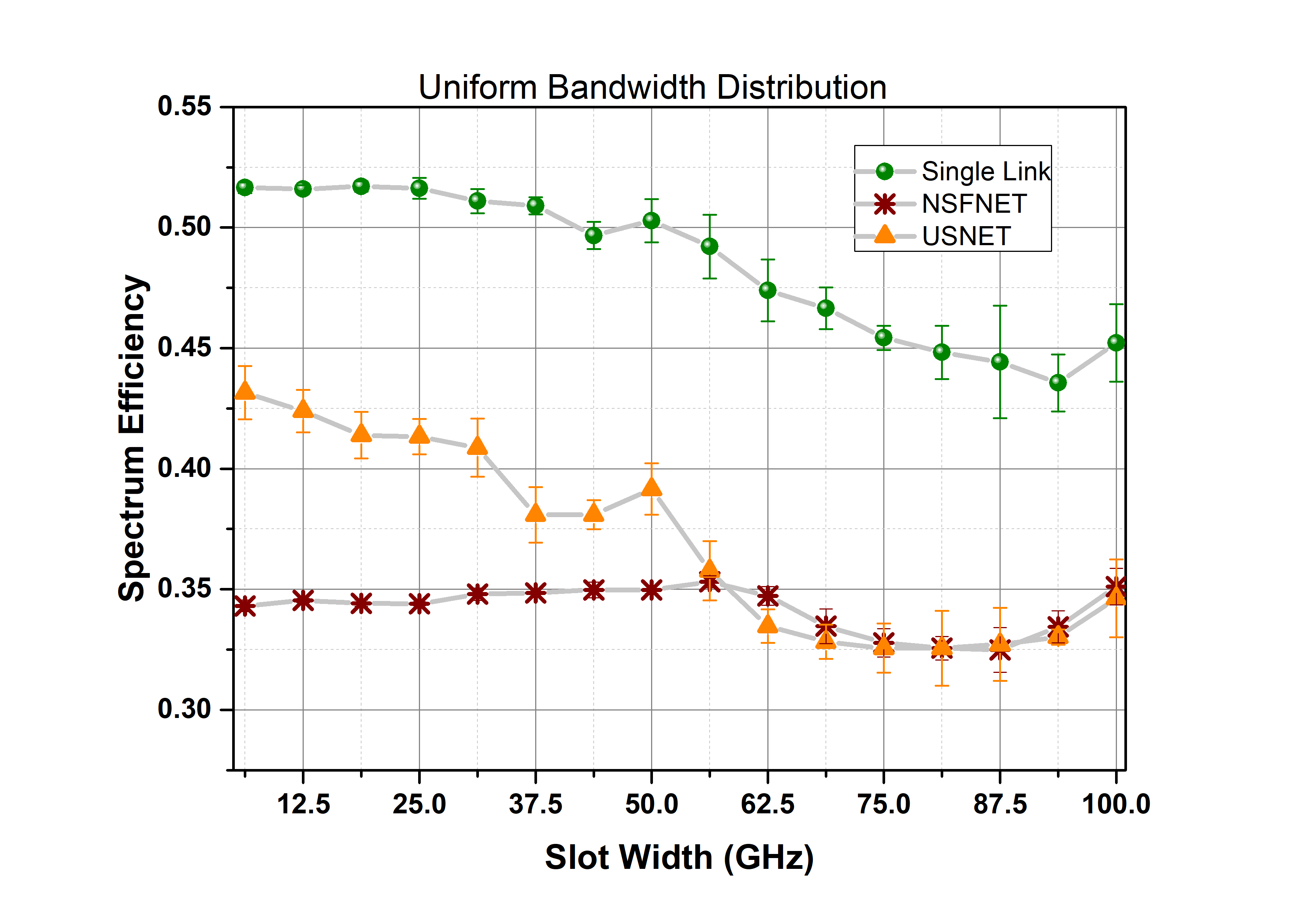}
         \caption{Spectrum Efficiency against various Slot Width.}
         \label{fig:seun}
     \end{subfigure}
        \caption{Various performance parameters for different network topologies, considering Offered Load per Node to be 20 Erlang and Average Bandwidth to be 100 Gbps}
        \label{fig:unin}
\end{figure}

\subsubsection{Uniform Bandwidth Distribution}
The bandwidth for all the lightpath requests on each node pair are uniformly distributed, i.e., the bandwidth required for each lightpath is chosen uniformly between $B_{min}$ and $B_{max}$. If only the $B_{max}$ is given, then in this paper, we assumed $B_{min}$ to be 1 Gbps. 

Figure \ref{fig:unin} shows plots of different performance parameters against various slot widths. In these plots, we have taken three network topologies as mentioned  in Section II-A. The blocking probability and bandwidth blocking probability for USNET are higher due to its higher nodal degree amongst the three topologies. In USNET, some of the lightpath requests can have higher path length. Higher path length means increase in the number of links. Hence, the requirement of continuous and contiguous slots also increases. Therefore, it results in higher fragmentation within the link spectra and thus higher blocking of the lightpath requests. In contrast, the spectrum efficiency of the single link is higher because there is only a single link connecting a node pair.  

In figure \ref{fig:bpun}, for the lower slot width, the blocking probability is low. Whereas as the slot width increases, the blocking probability also increases. The 6.75 GHz and 12.5 GHz work better for all the network topologies. Also, for NSFNET, there is almost zero blocking of lightpath requests for a slot width of 6.75 GHz. No point is shown in the plot since we cannot show zero on a logarithmic scale. A minor fall can be observed at other slot widths $\leq$50 GHz. The pattern of blocking probability is almost similar for all the network topologies. 

A similar plot can be seen in figure \ref{fig:bbpun} where the bandwidth blocking probability is compared for different network topologies for multiple slot widths. Here the conclusion is same as presented for blocking probability, except bandwidth blocking probability is slightly higher for all the slot sizes, because instead of counting the number of blocked connections, we are also taking into account the bandwidth of each lightpath request.

Another plot can be seen in figure \ref{fig:seun} where the spectrum efficiency is compared for different network topologies for multiple slot widths. The spectrum efficiency is higher when the blocking probability is low due to the less wastage of the assigned bandwidth. However, in the plot for NSFNET for lower slot widths, the spectrum efficiency is low. It is not due to bandwidth wastage but due to the under-utilization of spectrum slots. But as the network changes, the higher spectrum efficiency is also observed at same slot widths.  
\begin{figure}[h]
     \centering
     \captionsetup{justification=centering}

     \begin{subfigure}[b]{0.2\textwidth}
         \centering
         \includegraphics[width=\linewidth]{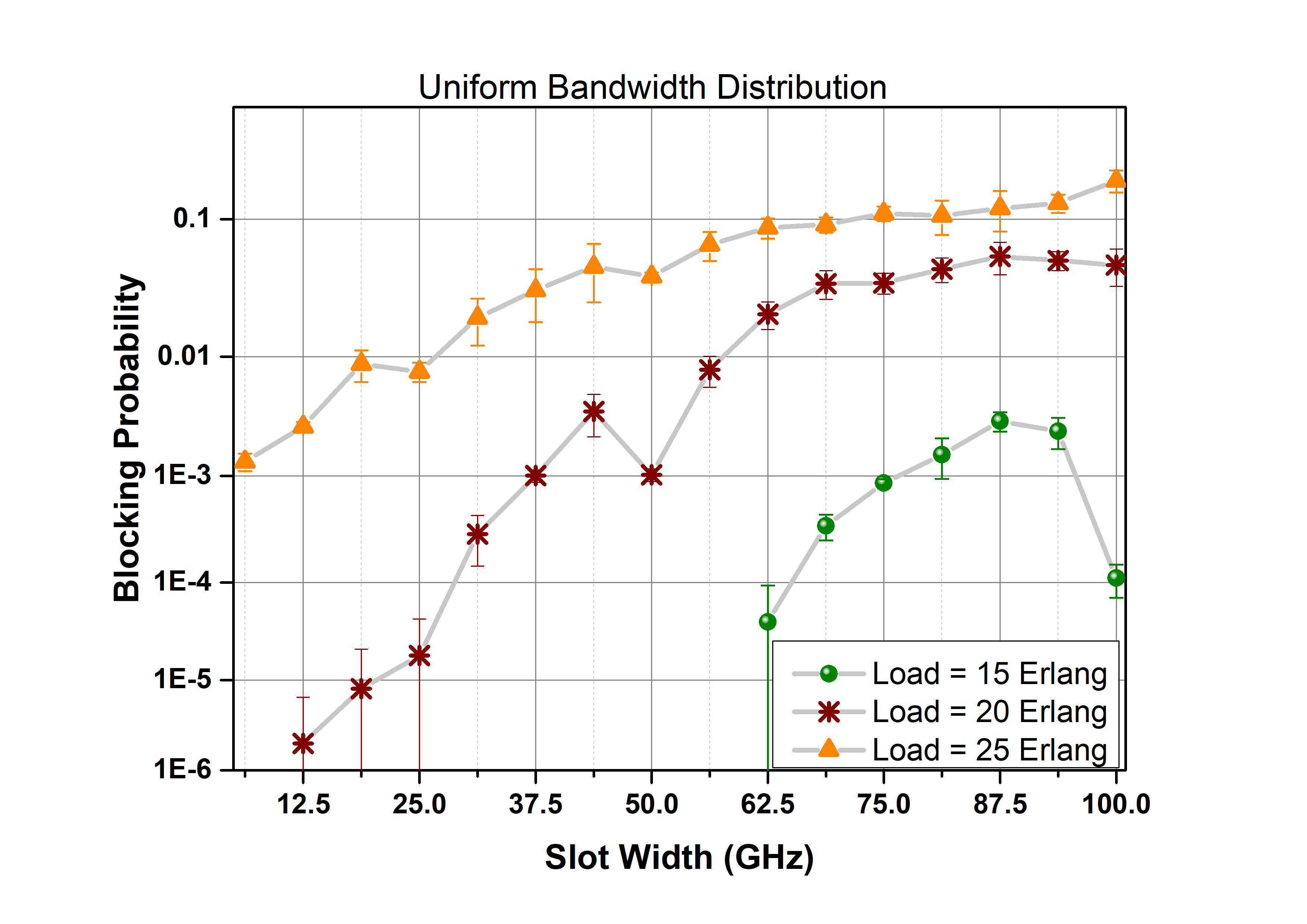}
         \caption{Blocking Probability against various Slot Width.}
         \label{fig:bpul}
     \end{subfigure}
     \hfill
     \begin{subfigure}[b]{0.2\textwidth}
         \centering
         \includegraphics[width=\linewidth]{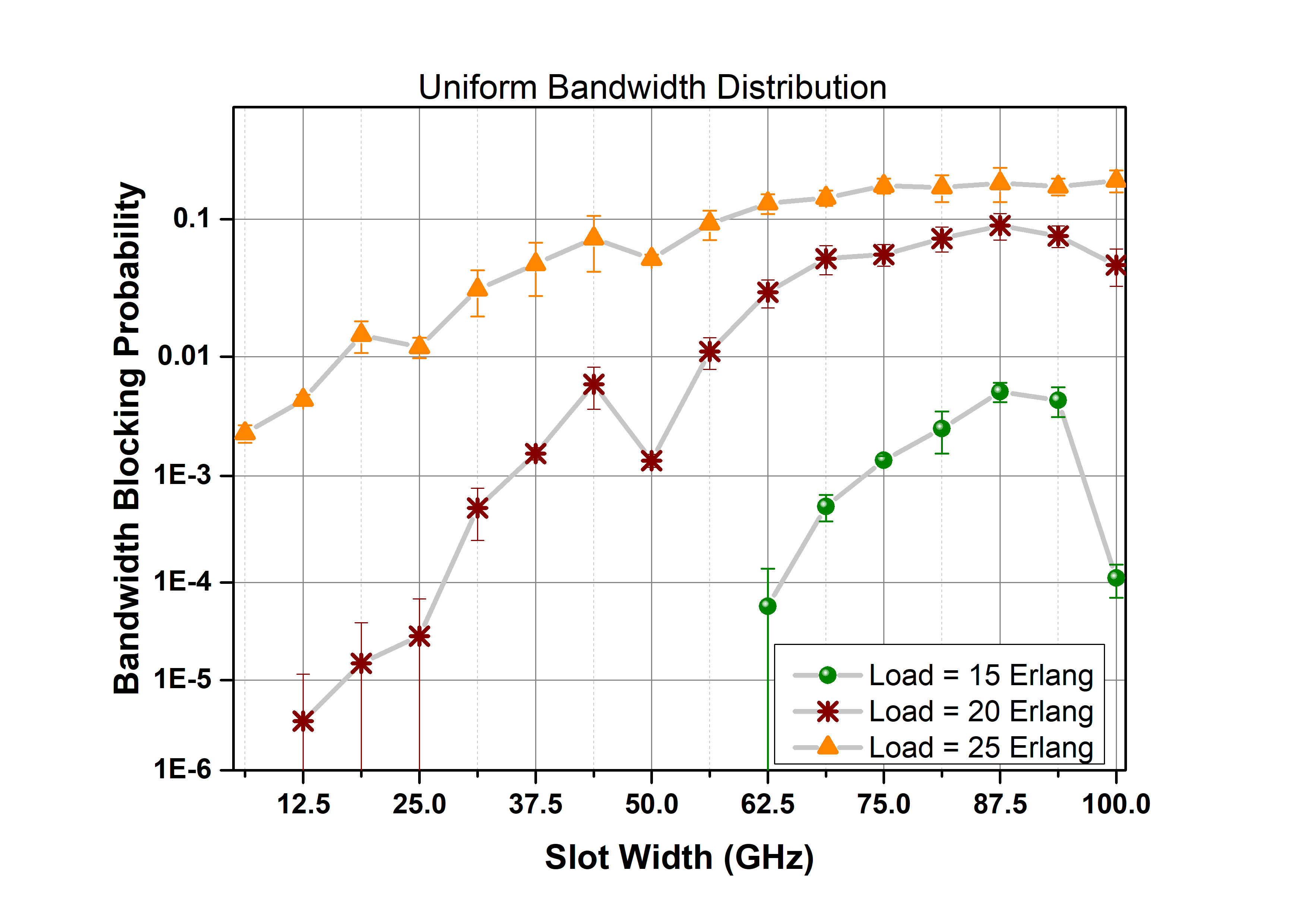}
         \caption{Bandwidth Blocking Probability against various Slot Width.}
         \label{fig:bbpul}
     \end{subfigure}
     \begin{subfigure}[b]{0.2\textwidth}
         \centering
         \includegraphics[width=\linewidth]{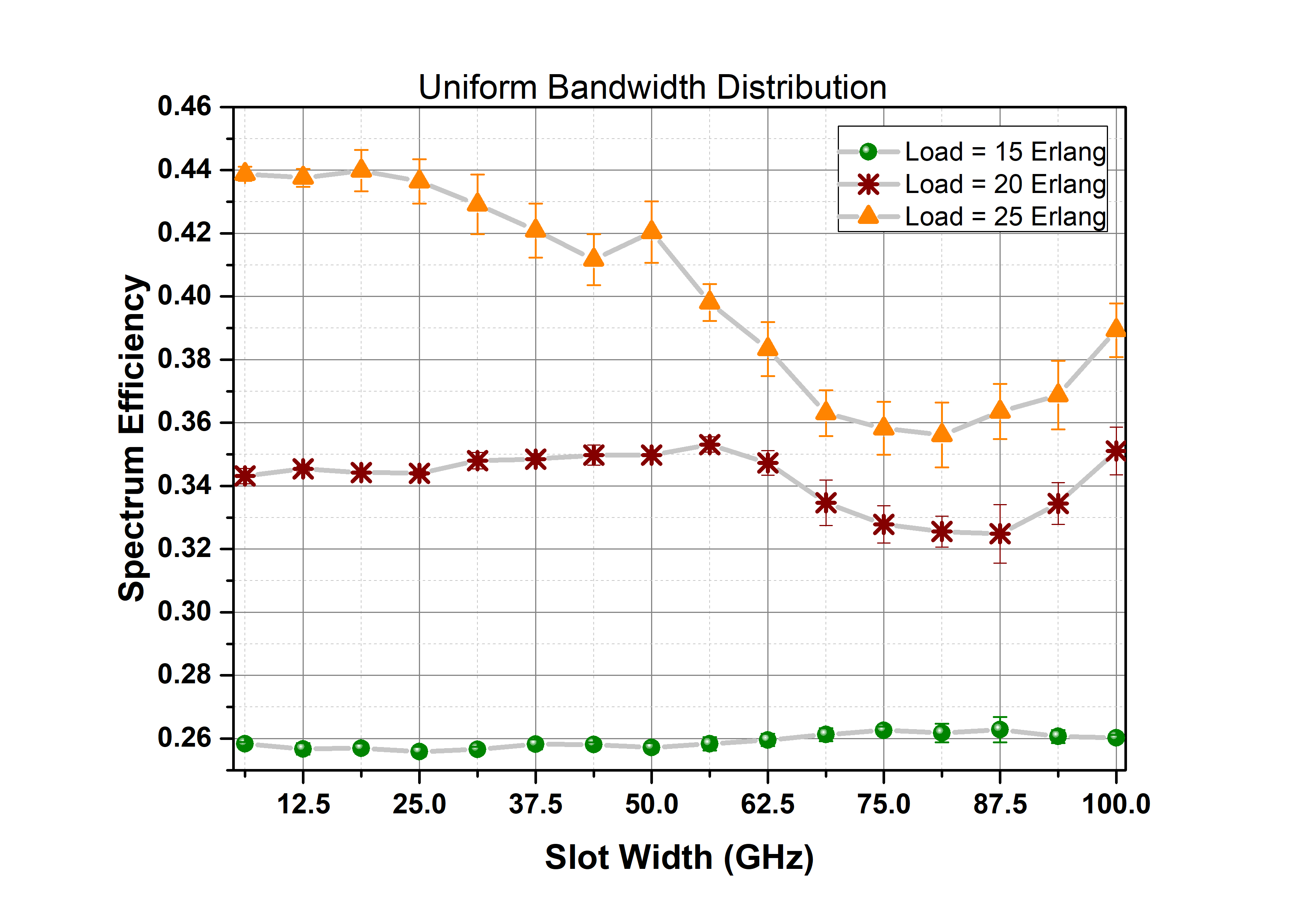}
         \caption{Spectrum Efficiency against various Slot Width.}
         \label{fig:seul}
     \end{subfigure}
        \caption{Various performance parameters for different Offered Load per Node, considering $B_{avg}$  to be 100 Gbps and NSFNET network topology.}
        \label{fig:unil}
\end{figure}

Figure \ref{fig:unil} shows plots of different performance parameters against various slot width. In these plots, we have taken different offered loads per node (in Erlang), considering $B_{max}$ = 100 Gbps and NSFNET Topology. The blocking probability, bandwidth blocking probability, and spectrum efficiency increases as the offered load per node increases. 

In figure \ref{fig:bpul}, for the lower slot width, the blocking probability is low. For offered load per node of 15 Erlang, the blocking of the lightpath request starts when the slot width is 62.5 GHz. No point is shown in the plot till 56.25 GHz since we cannot show zero on a logarithmic scale. Now, as the slot width increases, the blocking probability also increases. This rise is due to higher bandwidth wastage for higher slot width. Again there is a minor drop at 100 GHz. 

With the increase in the load, the blocking probability also increases. The difference in the maximum and minimum blocking probability decreases with the increase in the load. If we go for significantly higher loads, the blocking probability starts converging for all the slot widths.  

A similar plot can be seen in figure \ref{fig:bbpul} where the bandwidth blocking probability is compared for different offered loads per node, for different slot widths. Here the conclusion is same as presented for blocking probability, except bandwidth blocking probability is slightly higher for all the slot sizes, because instead of counting the number of blocked connections, we are taking into account the bandwidth of each lightpath request.

In figure \ref{fig:seul}, the spectrum efficiency versus slot width is compared for different offered loads per node. The spectrum efficiency is higher when the blocking probability is low due to the less wastage of the assigned bandwidth. 

The spectrum efficiency performance almost remains same with changing slot width for lower loads (15 Erlang). It is not due to bandwidth wastage but due to the under-utilization of spectrum slots. But as the load increases ($\geq$ 20 Erlang), the difference in the maximum and minimum spectrum efficiency increases. 

For higher loads, blocking probability range is small, thus slot width leading to maximum spectrum efficiency should be choosen. Whereas, for lower loads, the spectrum efficiency is almost same for lower slot widths, hence optimum slot width can be choosen based on the lowest blocking probability.

\begin{figure}[h]
     \centering
     \captionsetup{justification=centering}

     \begin{subfigure}[b]{0.2\textwidth}
         \centering
         \includegraphics[width=\linewidth]{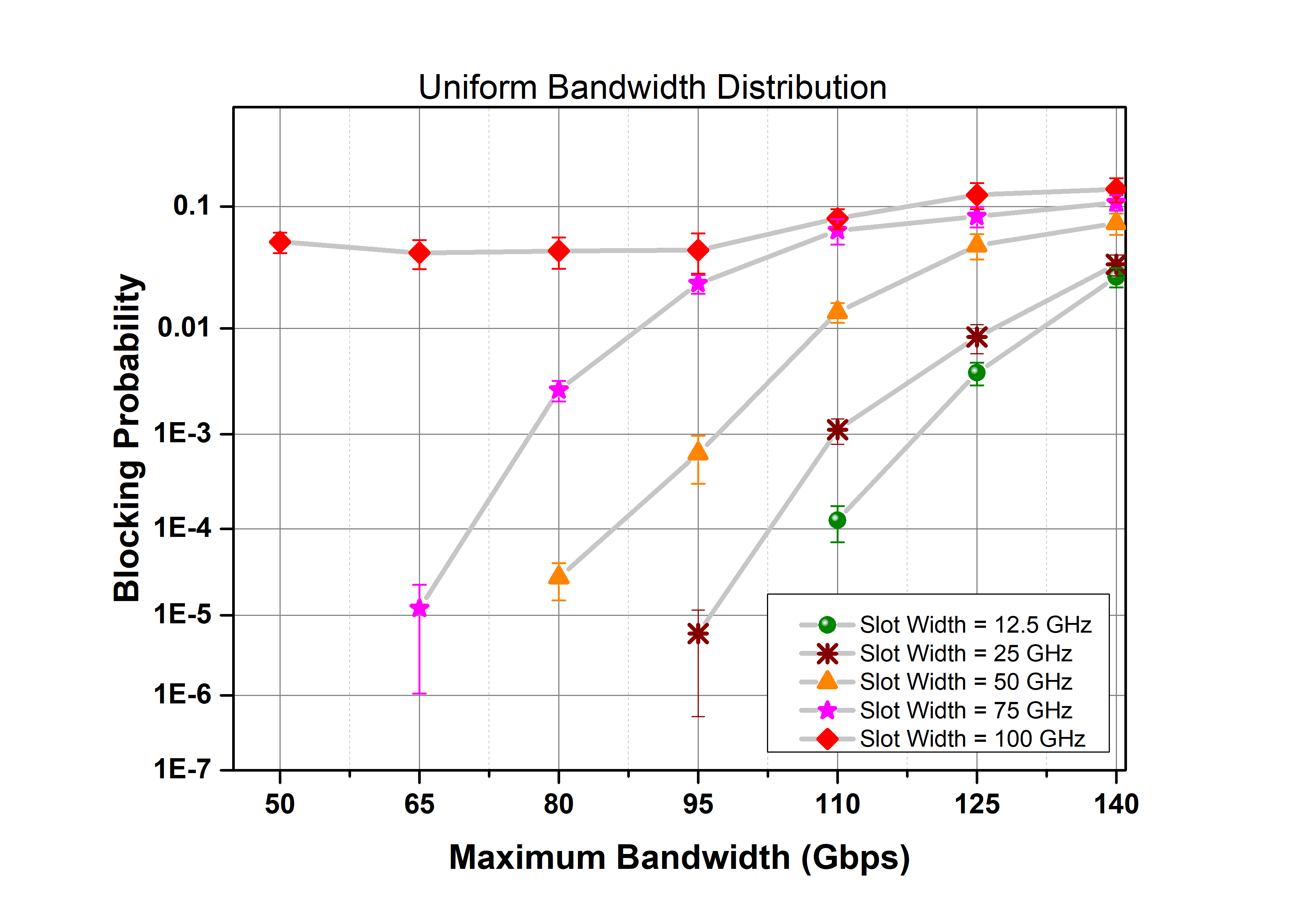}
         \caption{Blocking Probability against various Maximum Bandwidth $B_{max}$.}
         \label{fig:bpub}
     \end{subfigure}
     \hfill
     \begin{subfigure}[b]{0.2\textwidth}
         \centering
         \includegraphics[width=\linewidth]{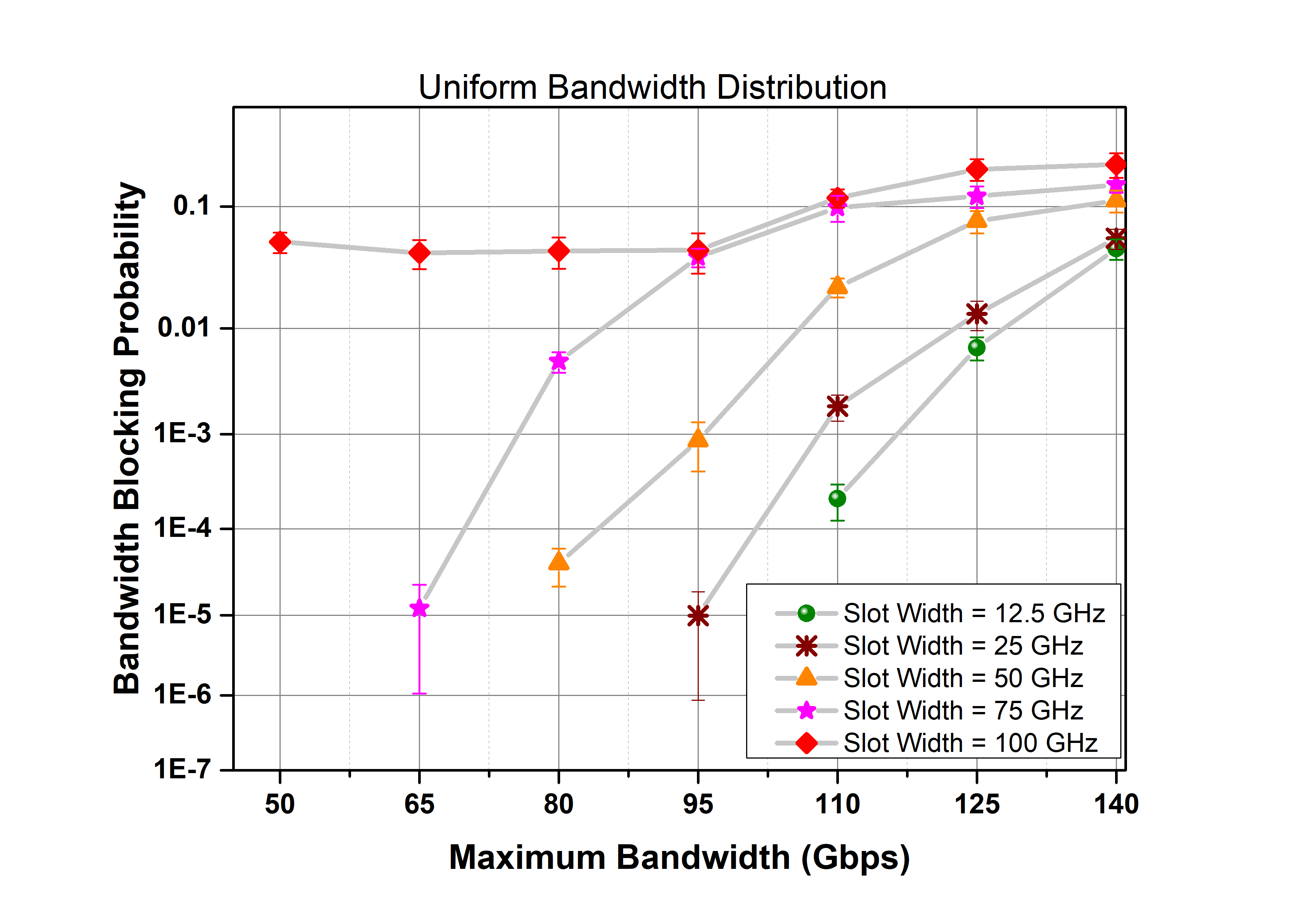}
         \caption{Bandwidth Blocking Probability against various Maximum Bandwidth $B_{max}$.}
         \label{fig:bbpub}
     \end{subfigure}
     \begin{subfigure}[b]{0.2\textwidth}
         \centering
         \includegraphics[width=\linewidth]{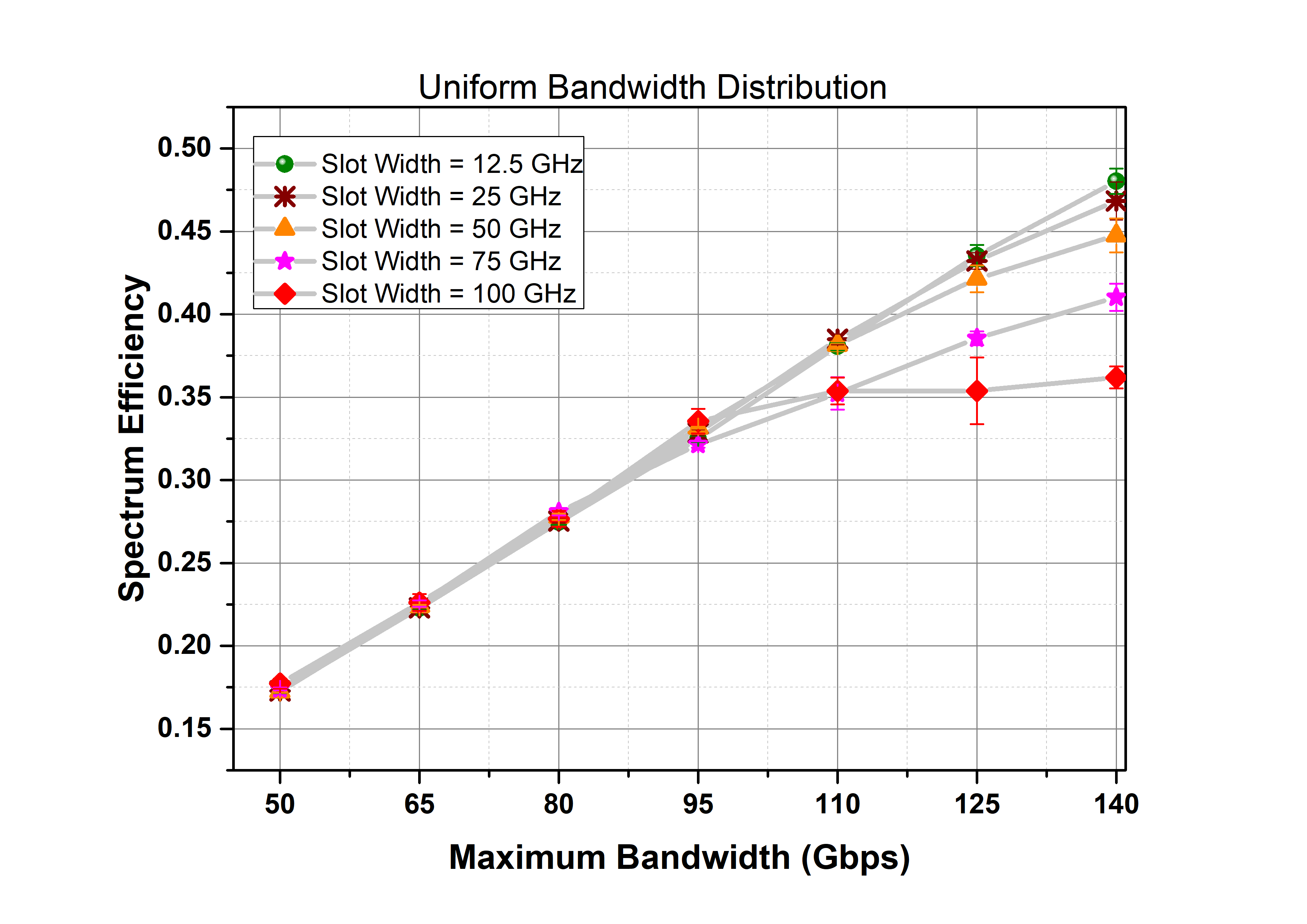}
         \caption{Spectrum Efficiency against various Maximum Bandwidth $B_{max}$.}
         \label{fig:seub}
     \end{subfigure}
        \caption{Various performance parameters for different slot widths, considering Offered Load per Node to be 20 Erlang and NSFNET network topology.}
        \label{fig:unib}
\end{figure}

Figure \ref{fig:unib} shows plots different performance parameters against different maximum bandwidth, $B_{max}$ (in Gbps). In these plots, we have taken different slot widths. The blocking probability and bandwidth blocking probability, and spectrum efficiency increase as the $B_{max}$ increases. Increase in $B_{max}$ implies increase in effective bandwidth load even for same load in Erlang. Erlang load only considers inter arrival time and connection duration, not the bandwidth requested by the connection.

In figure \ref{fig:bpub}, for the lower slot width, the blocking probability is low. We are considering the offered load per node to be 20 Erlang. No point is shown in the plot for lower values of $B_{max}$ because we cannot show zero on a logarithmic scale. Now, as the $B_{max}$ increases, the blocking probability also increases. The slot width of 12.5 GHz has the lowest blocking probability for all the $B_{max}$.  The blocking probability for all the slot widths converges for higher values of $B_{max}$.

A similar plot can be seen in figure \ref{fig:bbpub} where the bandwidth blocking probability is compared for multiple slot widths against various $B_{max}$. Here the conclusion is same as presented for blocking probability, except bandwidth blocking probability is slightly higher for all the slot sizes; because instead of counting the number of blocked connections, we are taking into account the bandwidth of each lightpath request.

An inverse plot can be seen in figure \ref{fig:seub} where the spectrum efficiency is compared for multiple slot widths against various $B_{max}$. The spectrum efficiency is higher when the blocking probability is low due to the minimum wastage of the assigned bandwidth.  The spectrum efficiency performance is almost same for the lower  $B_{max}$. It is not due to bandwidth wastage but due to the under-utilization of spectrum slots. But as the $B_{max}$ increases, the change in spectrum efficiency for various slot widths can be seen. The spectrum efficiency for 12.5 GHz is higher amongst all the considered slot widths for all the higher $B_{max}$, i.e., $>$ 95 Gbps.

Therefore, we can conclude for uniform bandwidth distribution, 6.25 GHz and 12.5 GHz outperform the other slot widths. Slot widths less than 6.25 or in between 6.25 GHz and 12.5 GHz can also have lower blocking probability and higher spectrum efficiency. But, here we are considering the slot widths to be $6.25 \times y$, where $y$ is a positive integer\footnote{6.25 GHz and 12.5 GHz are ITU-T slot widths.}. 

\subsubsection{Poisson Bandwidth Distribution}
Average bandwidth ($B_{avg}$ Gbps) characterize the bandwidth required by each lightpath request for the Poisson bandwidth distribution.

\begin{figure}[h]
     \centering
     \captionsetup{justification=centering}

     \begin{subfigure}[b]{0.2\textwidth}
         \centering
         \includegraphics[width=\linewidth]{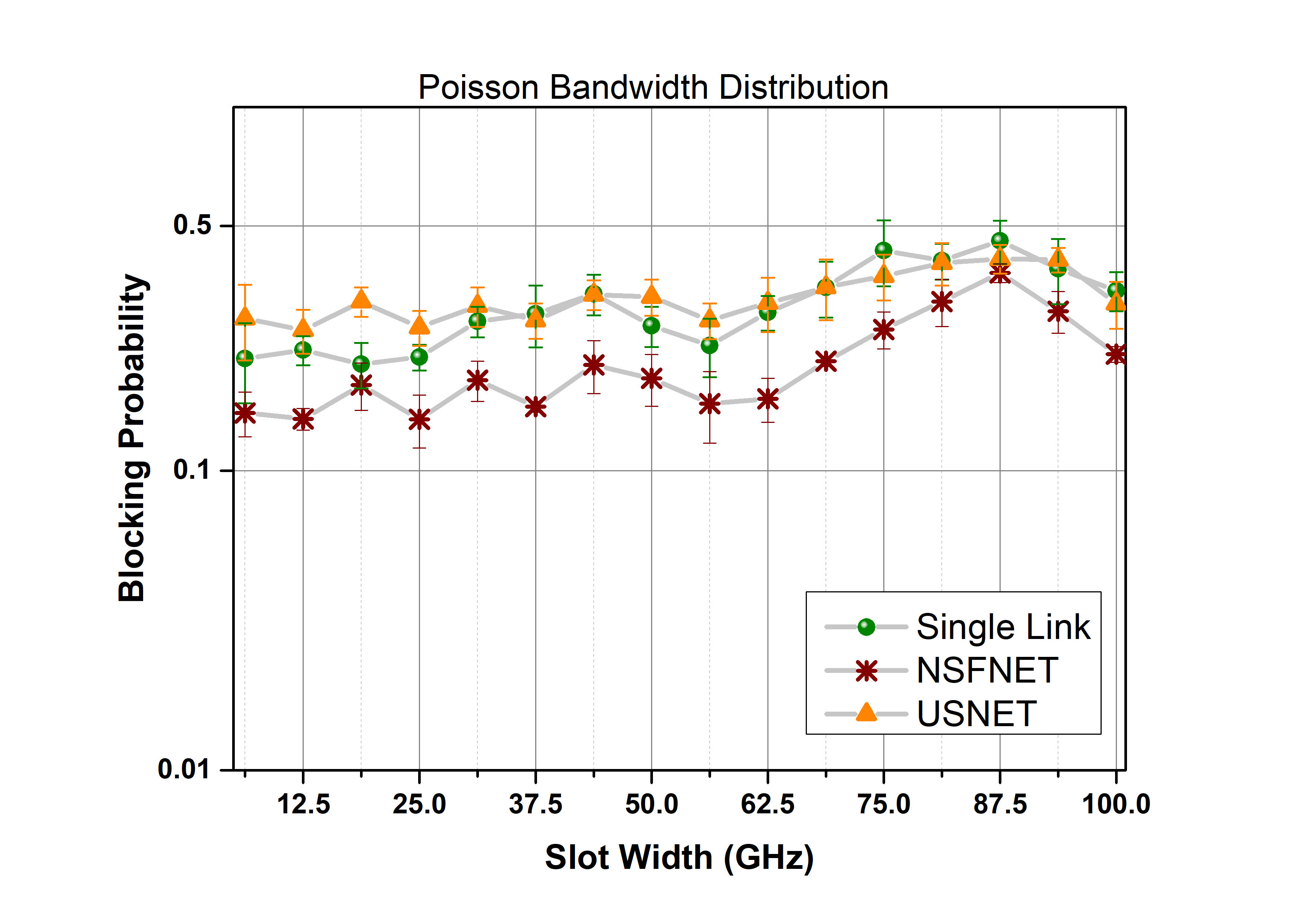}
         \caption{Blocking Probability against various Slot Widths.}
         \label{fig:bppn}
     \end{subfigure}
     \hfill
     \begin{subfigure}[b]{0.2\textwidth}
         \centering
         \includegraphics[width=\linewidth]{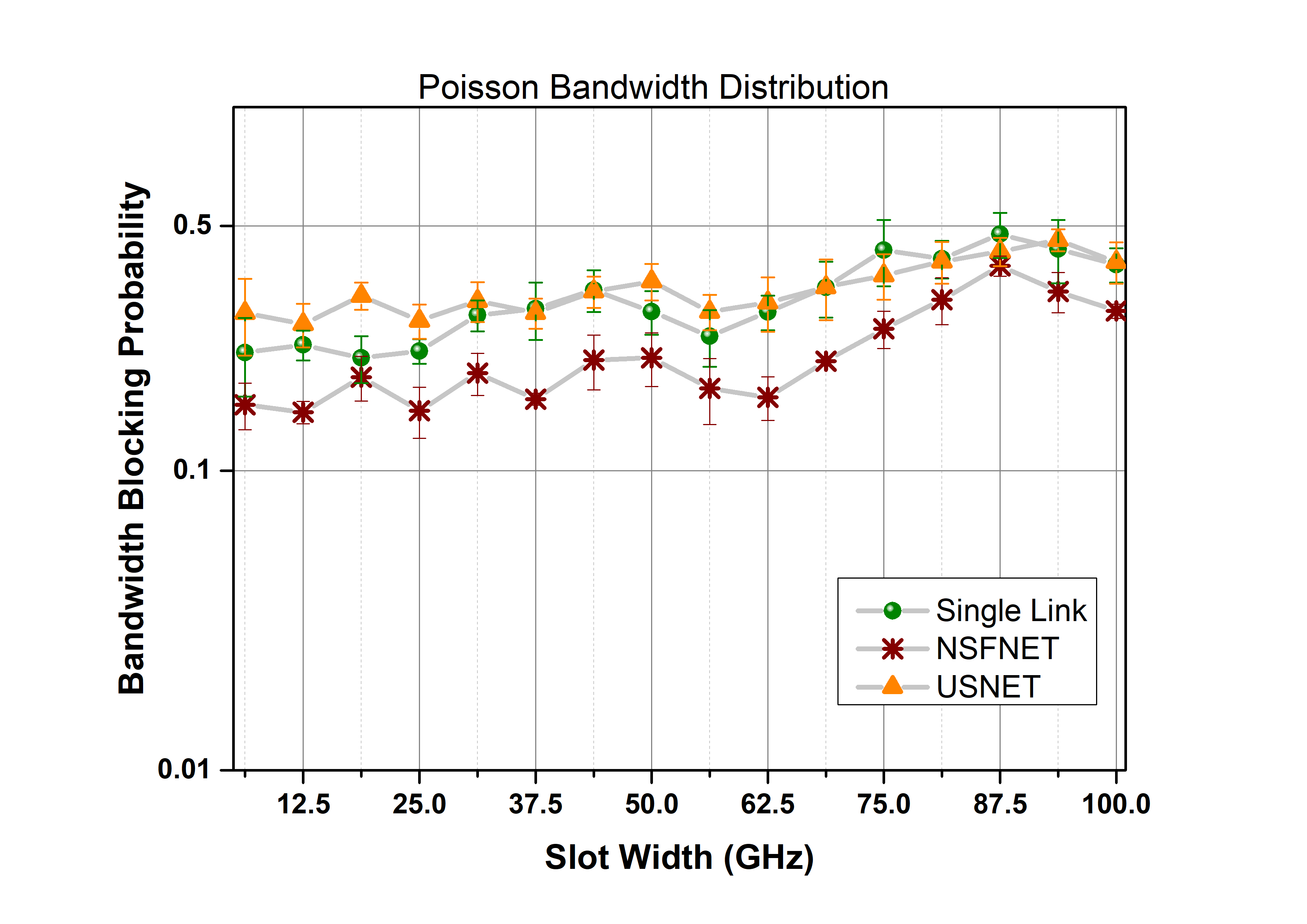}
         \caption{Bandwidth Blocking Probability against various Slot Widths.}
         \label{fig:bbppn}
     \end{subfigure}
     \begin{subfigure}[b]{0.2\textwidth}
         \centering
         \includegraphics[width=\linewidth]{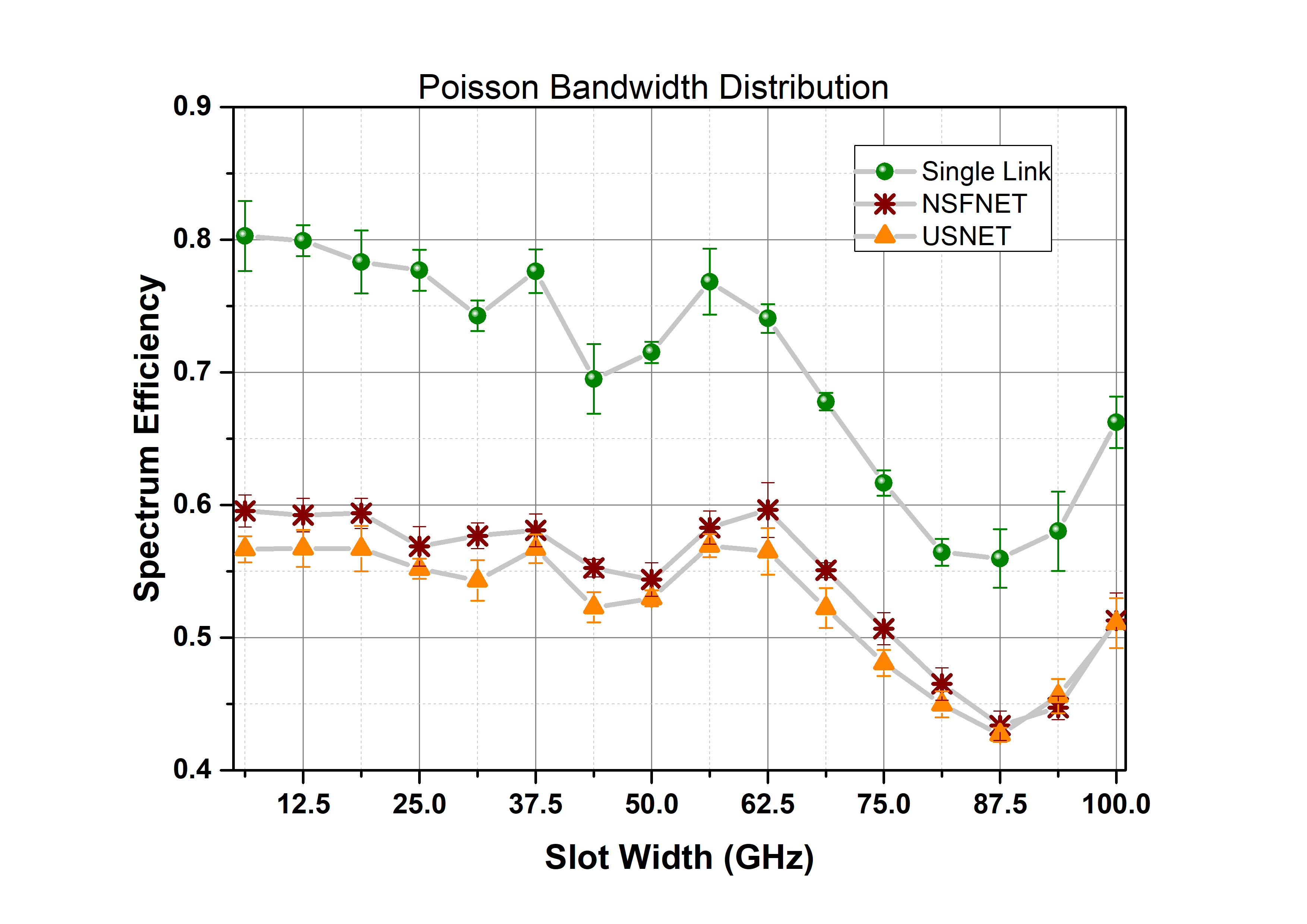}
         \caption{Spectrum Efficiency against various Slot Widths.}
         \label{fig:sepn}
     \end{subfigure}
        \caption{Various performance parameters for different network topologies, considering Offered Load per Node to be 20 Erlang and Average Bandwidth to be 100 Gbps}
        \label{fig:poin}
\end{figure}

Figure \ref{fig:poin} shows plots of different performance parameters against various slot widths for Poisson bandwidth distribution. In these plots, we have taken three network topologies, i.e., Single link, NSFNET, and USNET. 

Figure \ref{fig:bppn} shows the blocking probability is for Poisson bandwidth distribution. We can see for various slot widths, the blocking probability is lower. The slot width of sizes 6.25 GHz, 12.5 GHz, 25 GHz, 37.5 GHz, 56.25 GHz and 62.5 GHz have almost the same blocking probability value for NSFNET. The corresponding pattern can also be observed for USNET. Whereas for a single link, 6.25 GHz, 12.5 GHz, 18.75 GHz, 25 GHz, and 56.25 GHz have the lowest blocking probability. 

A similar plot can be seen in figure \ref{fig:bbppn}, where the bandwidth blocking probability is compared for different network topologies for various slot widths. Here the conclusions are the same as presented for blocking probability, except the bandwidth blocking probability is slightly higher for all the slot sizes.

The slot width of sizes 6.25 GHz, 12.5 GHz, 25 GHz, 37.5 GHz, 56.25 GHz and 62.5 GHz have a low blocking probability for NSFNET and USNET. We can further reduce this set by checking which slot size has maximum spectrum efficiency. We can use the same approach for a single link. 

Figure \ref{fig:sepn} shows the spectrum efficiency plot for Poisson bandwidth distribution against various slot width. The spectrum slot of 6.25 GHz and 12.5 GHz has maximum spectrum efficiency for all the network topologies. This means less bandwidth wastage.

\begin{figure}[h]
     \centering
     \captionsetup{justification=centering}

     \begin{subfigure}[b]{0.2\textwidth}
         \centering
         \includegraphics[width=\linewidth]{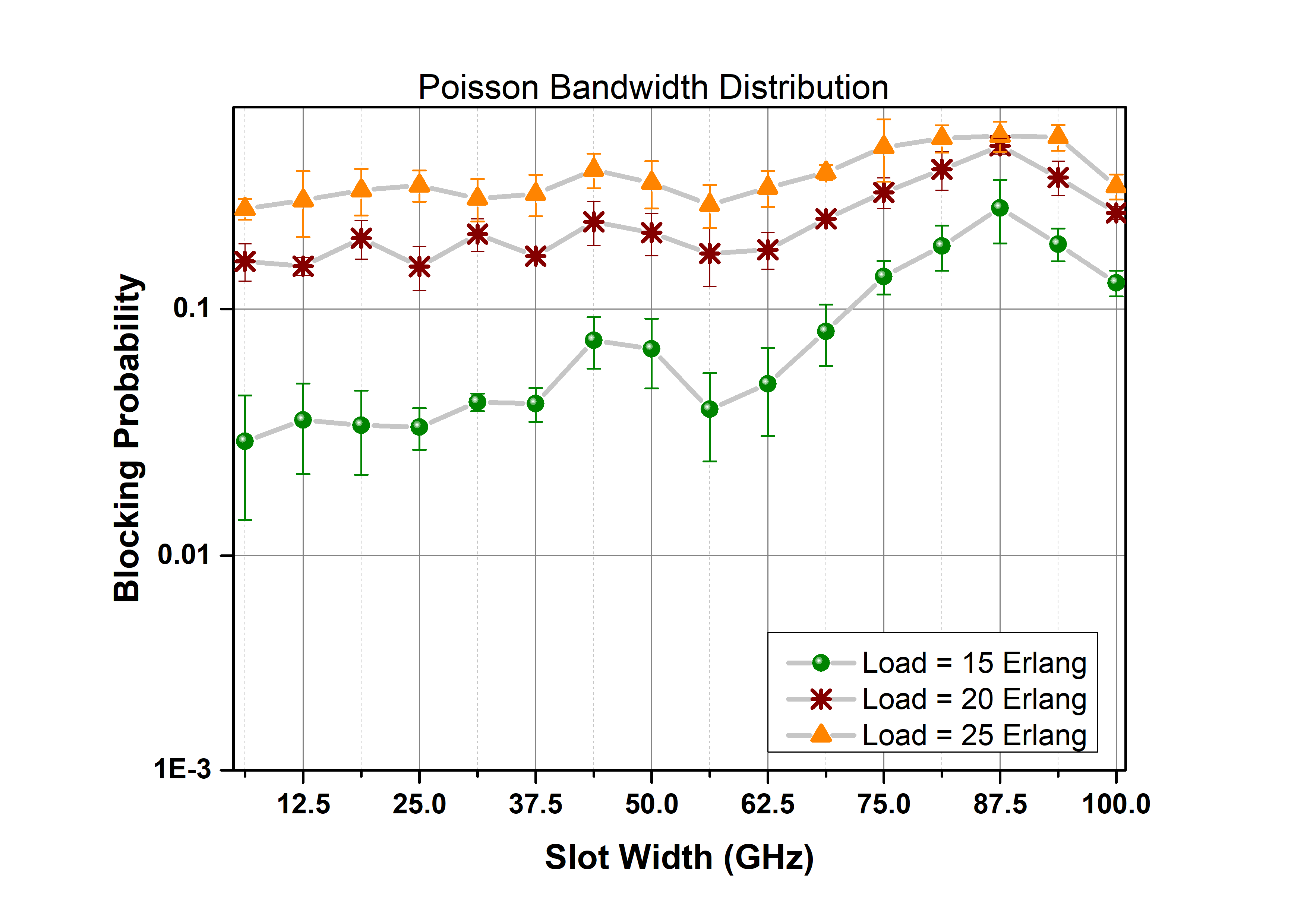}
         \caption{Blocking Probability against various Slot Width.}
         \label{fig:bppl}
     \end{subfigure}
     \hfill
     \begin{subfigure}[b]{0.2\textwidth}
         \centering
         \includegraphics[width=\linewidth]{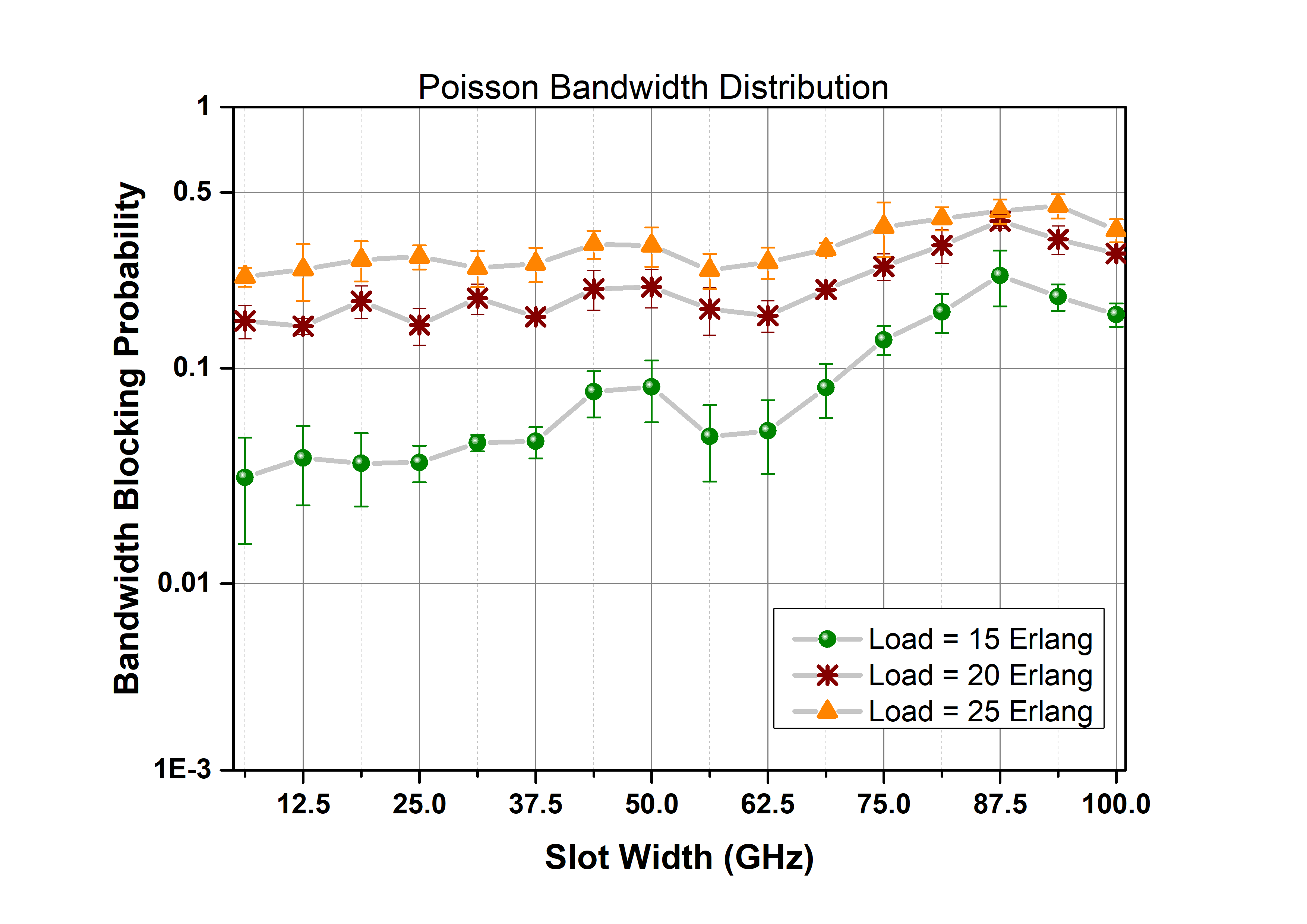}
         \caption{Bandwidth Blocking Probability against various Slot Width.}
         \label{fig:bbppl}
     \end{subfigure}
     \begin{subfigure}[b]{0.2\textwidth}
         \centering
         \includegraphics[width=\linewidth]{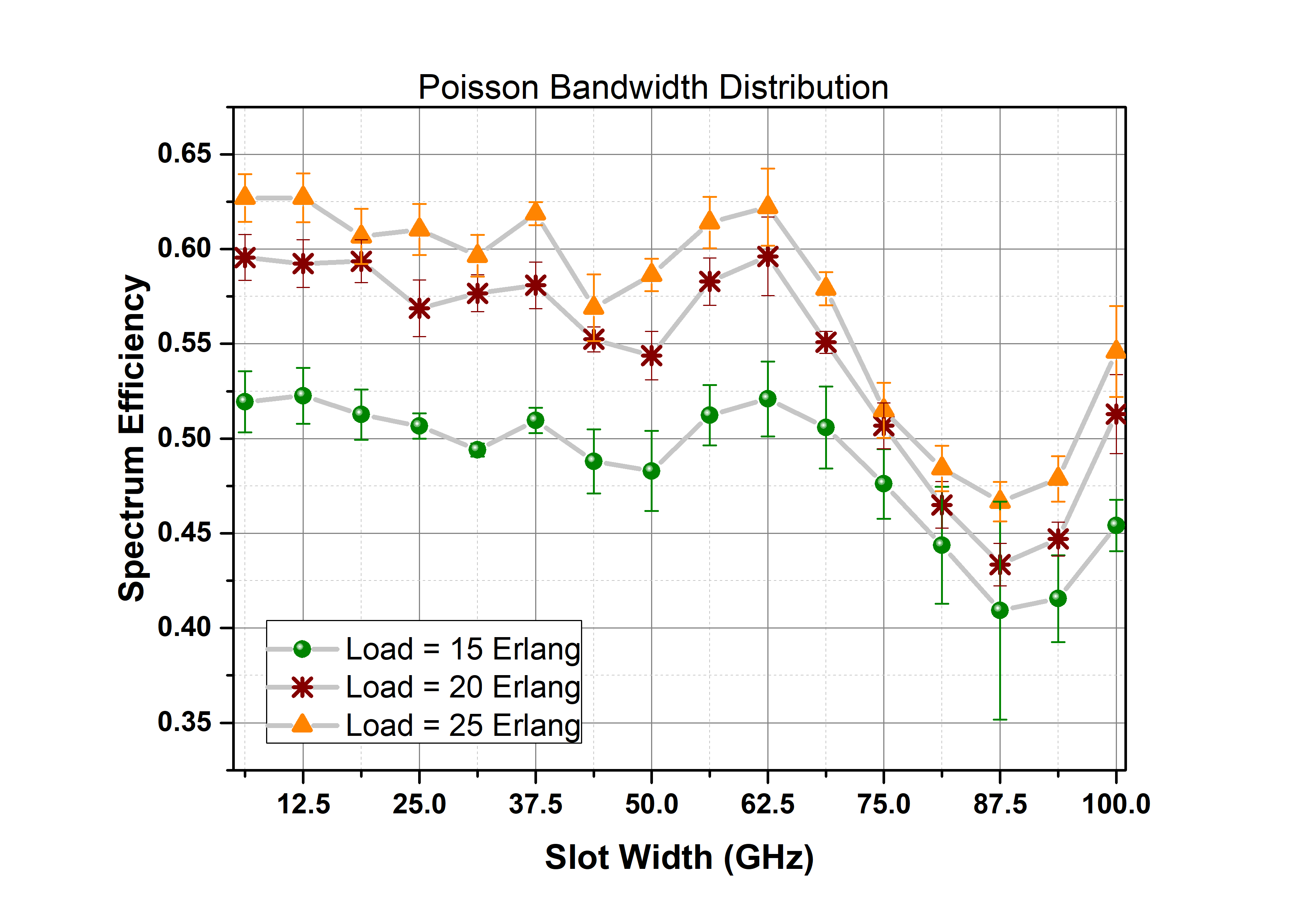}
         \caption{Spectrum Efficiency against various Slot Width.}
         \label{fig:sepl}
     \end{subfigure}
        \caption{Various performance parameters for different Offered Load per Node, considering $B_{avg}$  to be 100 Gbps and NSFNET network topology.}
        \label{fig:poil}
\end{figure}

Figure \ref{fig:poil} shows  plots of different performance parameters against various slot widths.  In these plots, we have taken different offered loads per node (in Erlang). The blocking probability, bandwidth blocking probability, and spectrum efficiency increase as the offered load per node increases. 

Figure \ref{fig:bppl} shows the blocking probability is for Poisson bandwidth distribution. We can see for various slot widths, the blocking probability is lower. Blocking probability for the slot widths of sizes 6.25 GHz, 12.5 GHz, 25 GHz, 37.5 GHz, and 56.25 GHz, are almost same for the offered loads per node values of 20 and 25 Erlangs. For 15 Erlang load blocking probabilities tends to increase with increasing slot width.

A similar plot can be seen in figure \ref{fig:bbppl} where the bandwidth blocking probability is compared for different network topologies for multiple slot widths. Here, also conclusions are same as that for blocking probability, except bandwidth blocking probability is slightly higher for all the slot sizes; because instead of counting the number of blocked connections, we are taking into account the bandwidth of each lightpath request.

The slot width of sizes 6.25 GHz, 12.5 GHz, 25 GHz, 37.5 GHz, and 56.25 GHz have a low blocking probability for all the offered loads per node.  We can further reduce this set by checking which slot sizes have maximum spectrum efficiency. 

Figure \ref{fig:sepl} shows the spectrum efficiency plot for Poisson bandwidth distribution against various slot widths. The spectrum slot width of 6.25 GHz and 12.5 GHz has maximum spectrum efficiency for all the network topologies. This means less bandwidth wastage.  

\begin{figure}[h]
     \centering
     \captionsetup{justification=centering}

     \begin{subfigure}[b]{0.2\textwidth}
         \centering
         \includegraphics[width=\linewidth]{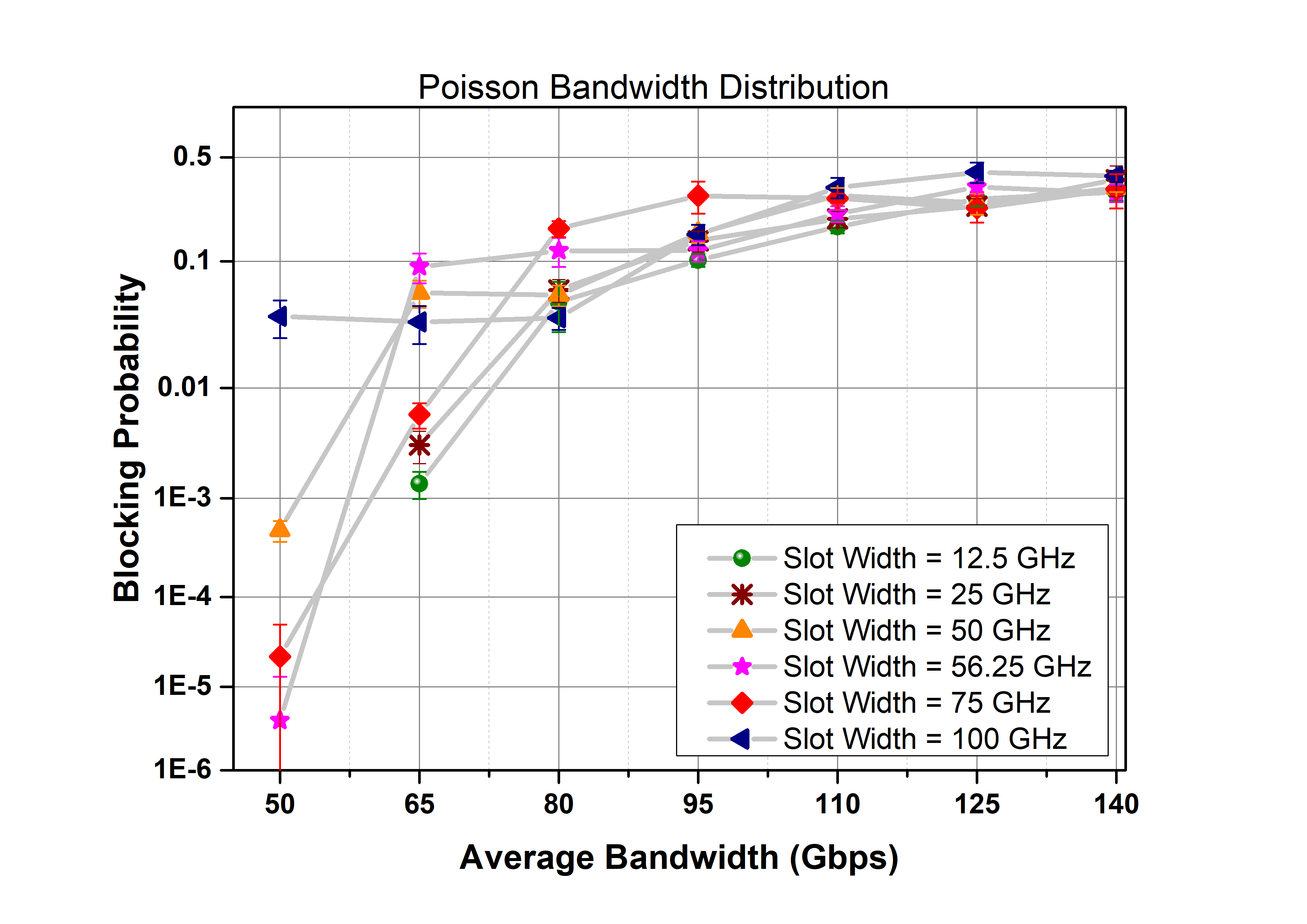}
         \caption{Blocking Probability against various Average Bandwidth.}
         \label{fig:bppb}
     \end{subfigure}
     \hfill
     \begin{subfigure}[b]{0.2\textwidth}
         \centering
         \includegraphics[width=\linewidth]{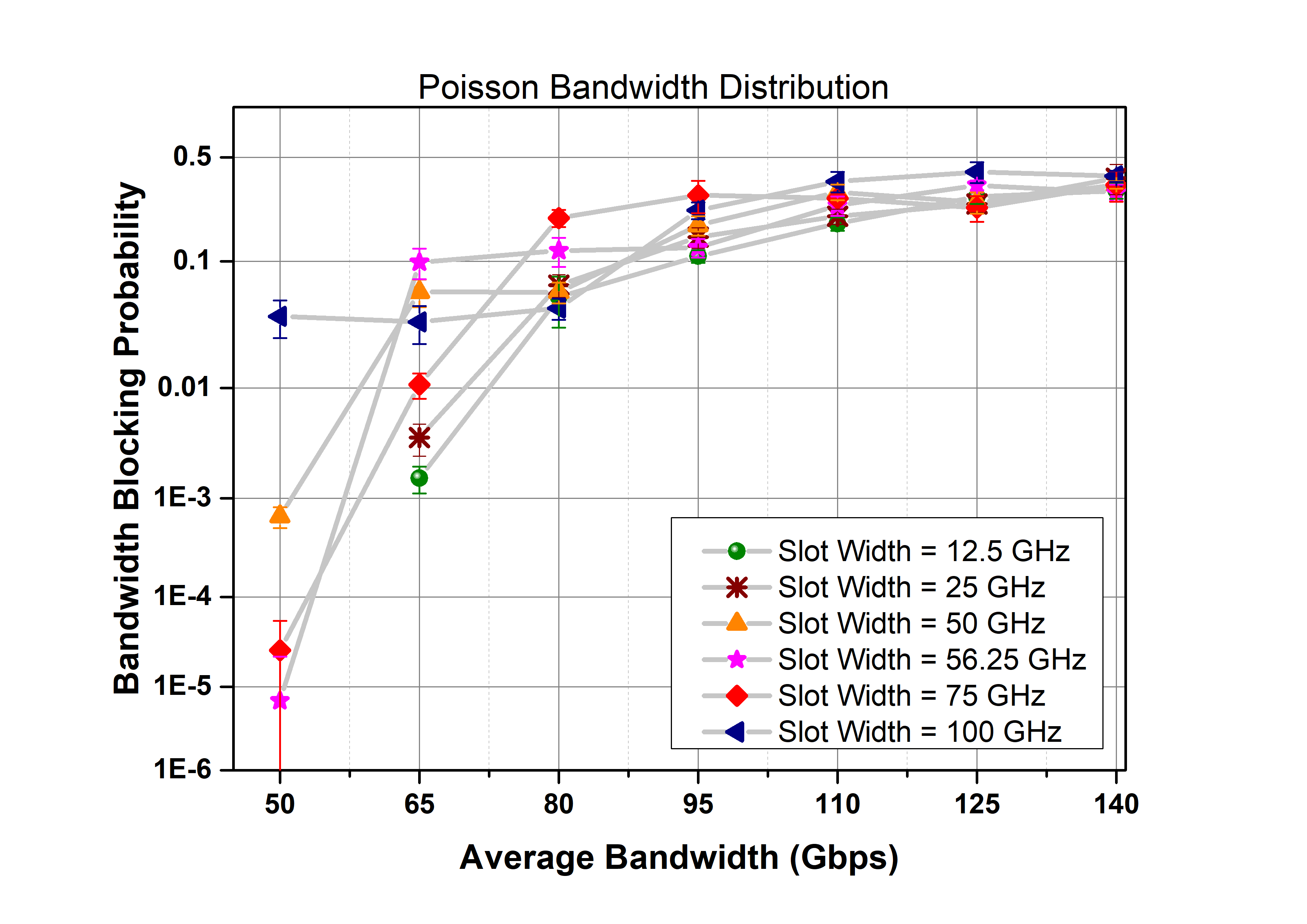}
         \caption{Bandwidth Blocking Probability against various Average Bandwidth.}
         \label{fig:bbppb}
     \end{subfigure}
     \begin{subfigure}[b]{0.2\textwidth}
         \centering
         \includegraphics[width=\linewidth]{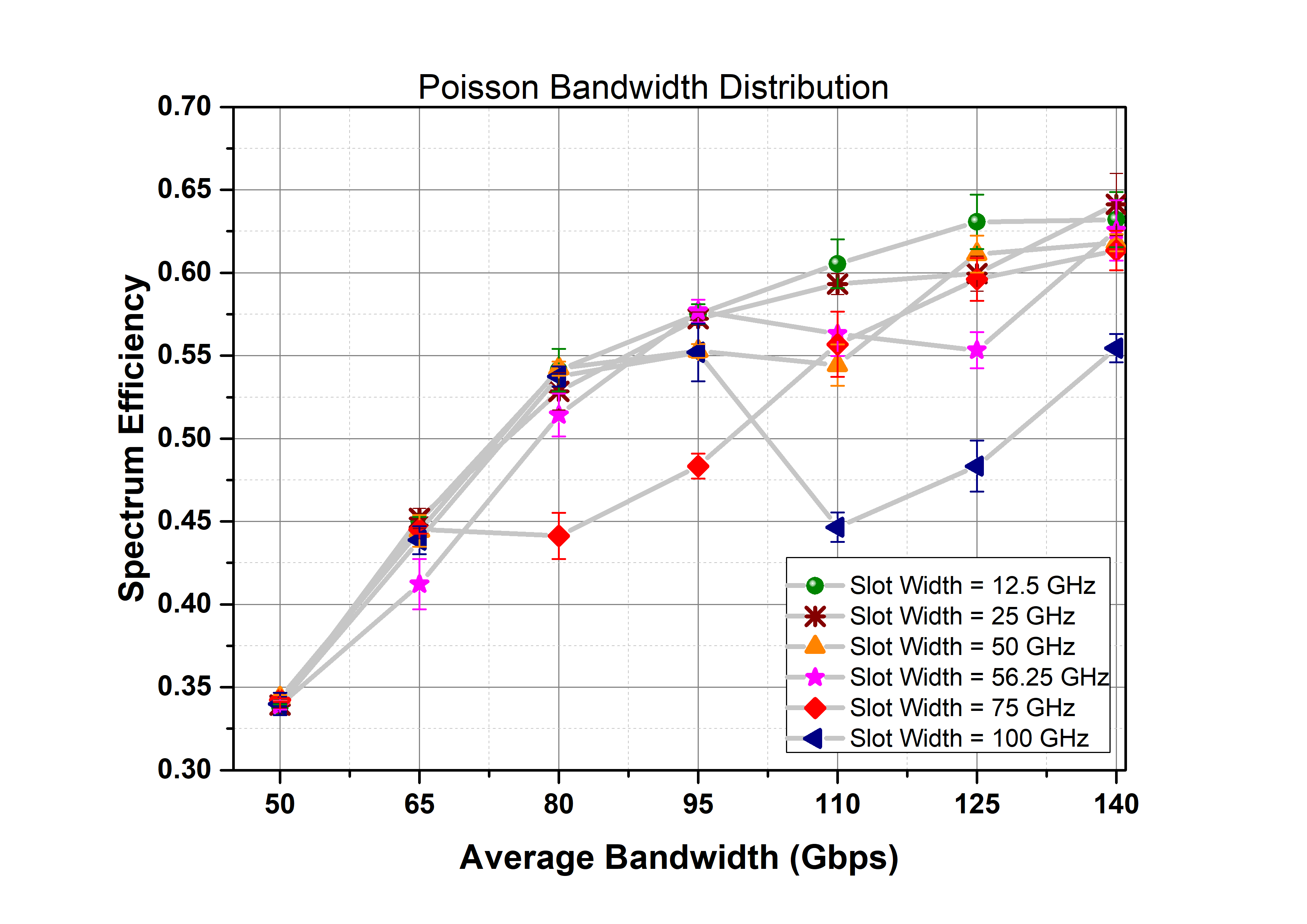}
         \caption{Spectrum Efficiency against various Average Bandwidth.}
         \label{fig:sepb}
     \end{subfigure}
        \caption{Various performance parameters for different slot widths, considering Offered Load per Node to be 20 Erlang for NSFNET network topology.}
        \label{fig:poib}
\end{figure}

In figures \ref{fig:poin} and \ref{fig:poil}, 56.25 GHz is also one of the near optimal slot width for $B_{avg}$ of 100 Gbps. However, this slot width performance might change with the $B_{avg}$ values. The performance various slot widths can be seen in the next set of plots (Figure \ref{fig:poib}).   

Figure \ref{fig:poib} shows the plots of different performance parameters against various average bandwidth, $B_{avg}$ (in Gbps). In these plots, we have taken different slot widths. Generally, the blocking probability and bandwidth blocking probability, and spectrum efficiency increase as the $B_{avg}$ increases. 

In figure \ref{fig:bppb}, for the lower slot width, the blocking probability is low. We are considering the offered load per node of 20 Erlangs. As the $B_{avg}$ increases, the blocking probability also increases. The slot width of 12.5 GHz has the lowest blocking probability for all the $B_{avg}$.  The blocking probability of all the slot widths converges for higher values of $B_{avg}$.

A similar plot can be seen in figure \ref{fig:bbppb} where the bandwidth blocking probability is compared for multiple slot widths against various $B_{avg}$. Here the conclusion is the same as presented for blocking probability, except bandwidth blocking probability is slightly higher for all the slot sizes. 

In figure \ref{fig:sepb}, spectrum efficiency is compared for multiple slot widths against various $B_{avg}$. The spectrum efficiency is higher when the blocking probability is low due to the less wastage of the assigned bandwidth.  The spectrum efficiency performance is almost same for $B_{avg}$ = 50 Gbps. It will also be there for $B_{avg} <$ 50 Gbps  It is not due to bandwidth wastage but due to the under-utilization of spectrum slots. But as the $B_{avg}$ increases, the change in spectrum efficiency for various slot widths can be seen. The spectrum efficiency for 12.5 GHz is highest amongst all the considered slot widths. For some of the slot widths, spectrum efficiency drop when $B_{avg}$ is around multiple of slot width.

In figures \ref{fig:poin} and \ref{fig:poil}, we have seen slot width of the size of 56.25 GHz performing better as compared to other higher slot widths. In figure \ref{fig:poib}, the performance parameters for 56.25 GHz are plotted against various $B_{avg}$. Here, we can see that for not all $B_{avg}$, 56.25 GHz performs near-optimally as 12.5 GHz does.

For Poisson bandwidth distribution also, 6.25 GHz and 12.5 GHz outperform the other slot widths. Slot widths less than 6.25 or in between 6.25 GHz and 12.5 GHz can also have lower blocking probability and higher spectrum efficiency. But, here we are considering the slot widths $6.25 \times y$, where $y$ is a positive integer\footnote{6.25 GHz and 12.5 GHz are ITU-T slot widths.}. 

\subsubsection{Constant Bandwidth}
The bandwidth for all the lightpath requests on each node pair is constant, i.e., $B$ Gbps. 

\begin{figure}[h]
     \centering
     \captionsetup{justification=centering}

     \begin{subfigure}[b]{0.2\textwidth}
         \centering
         \includegraphics[width=\linewidth]{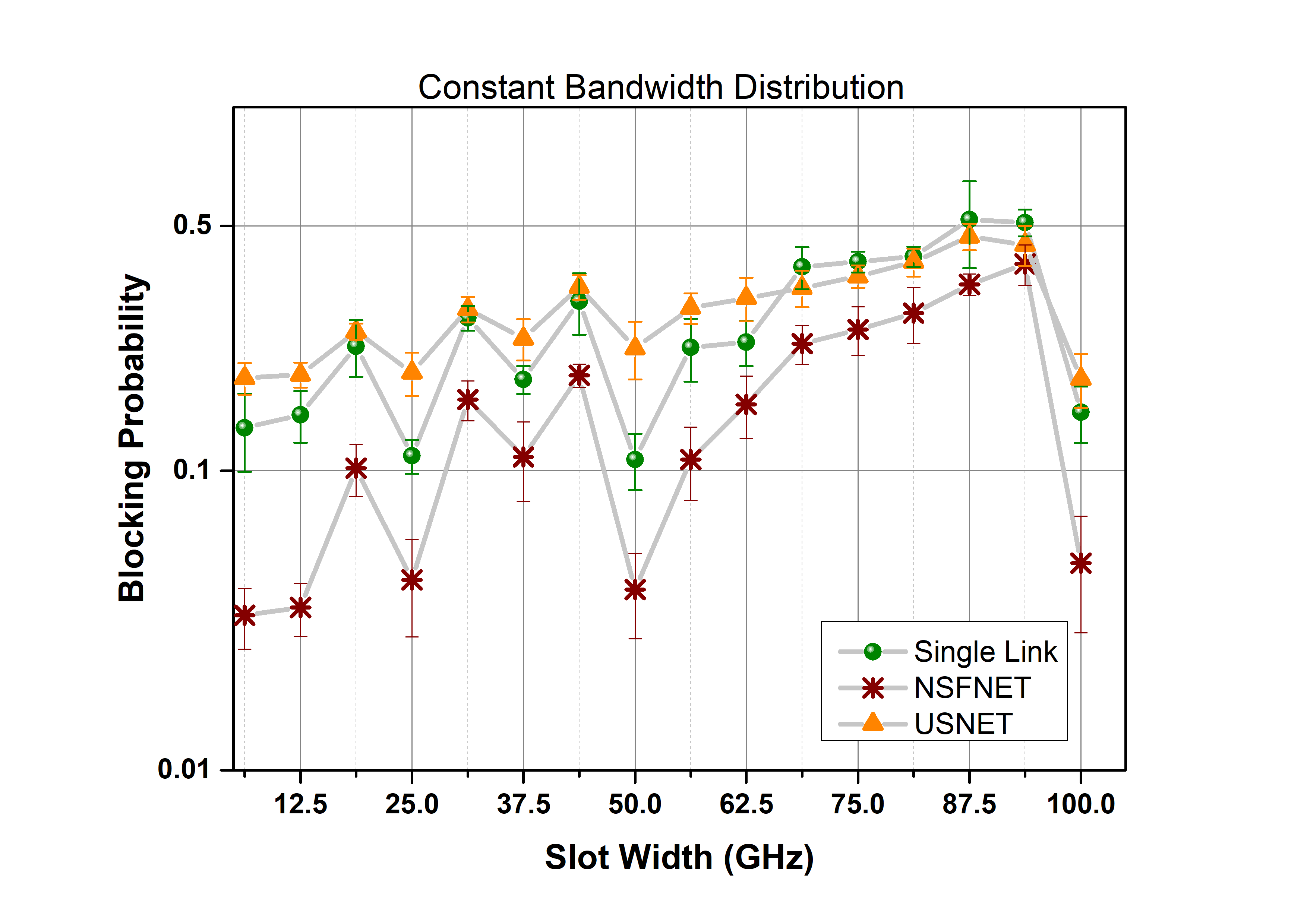}
         \caption{Blocking Probability against various Slot Width.}
         \label{fig:bpcn}
     \end{subfigure}
     \hfill
     \begin{subfigure}[b]{0.2\textwidth}
         \centering
         \includegraphics[width=\linewidth]{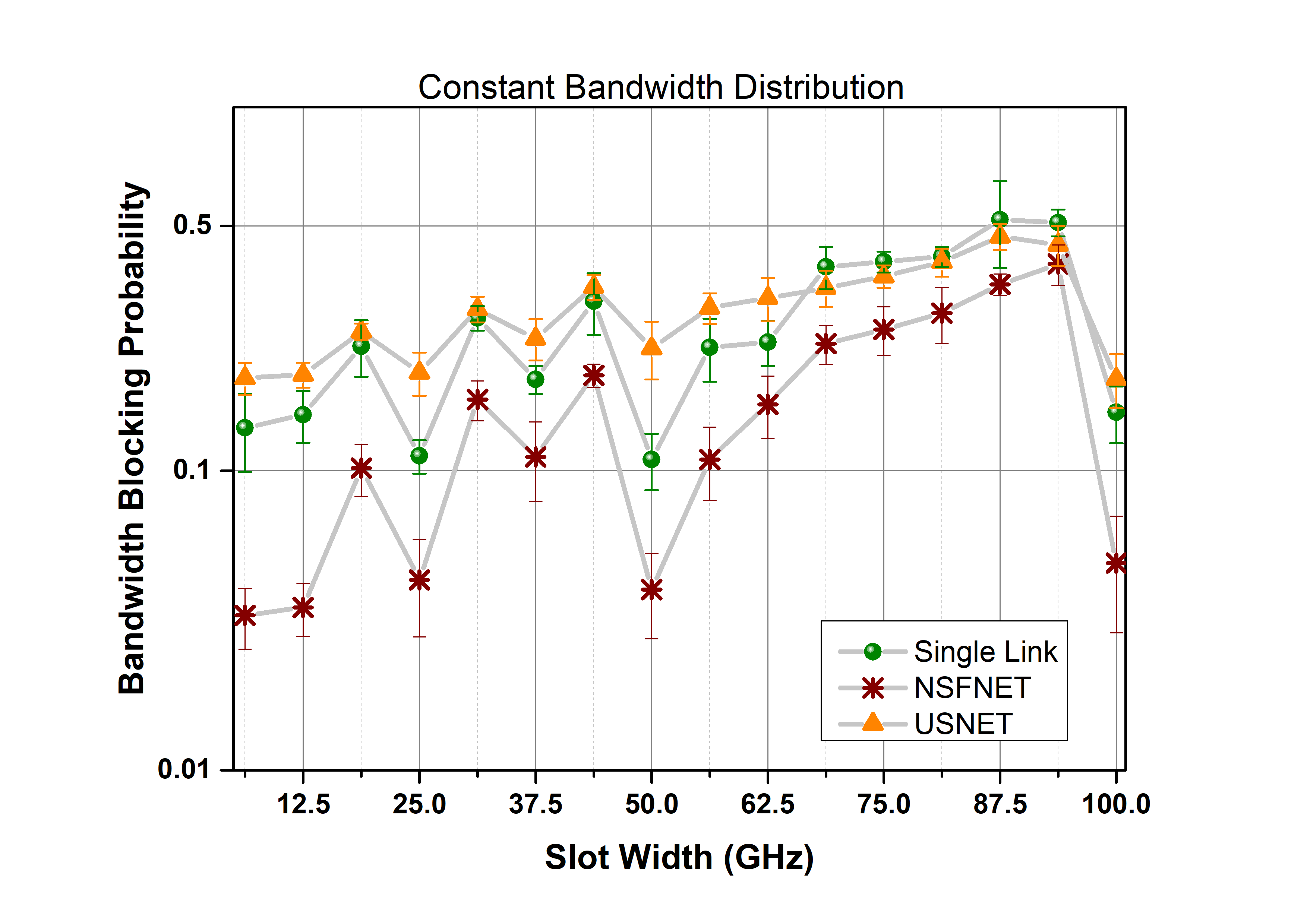}
         \caption{Bandwidth Blocking Probability against various Slot Width.}
         \label{fig:bbpcn}
     \end{subfigure}
     \begin{subfigure}[b]{0.2\textwidth}
         \centering
         \includegraphics[width=\linewidth]{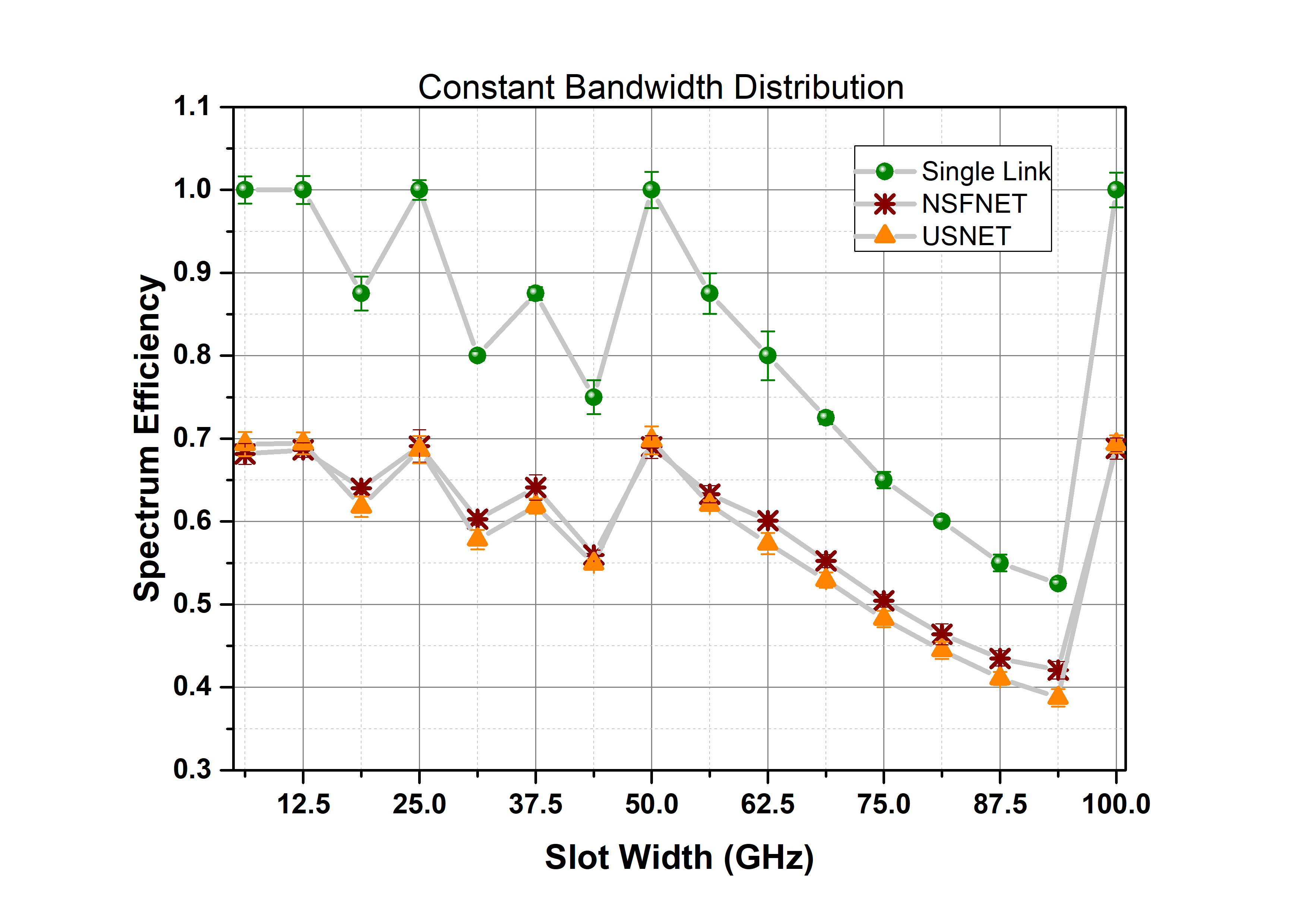}
         \caption{Spectrum Efficiency against various Slot Width.}
         \label{fig:secn}
     \end{subfigure}
        \caption{Various performance parameters for different network topologies, considering Offered Load per Node to be 20 Erlang and Bandwidth to be 100 Gbps}
        \label{fig:consn}
\end{figure}

Figure \ref{fig:consn} shows plots of different performance parameters against various slot widths. In these plots, we have taken three network topologies.

Figure \ref{fig:bpcn} shows the blocking probability for constant traffic against various slot sizes. There are dips in the blocking probability at a few of the points in the plot. The slot width 6.25 GHz, 12.5 GHz, 25 GHz, 50 GHz and 100 GHz has a minima as the network size increases. 100 GHz performs better than other slot sizes because there is no impact of contiguity constraint; therefore, the chances of fragmentation reduce significantly.

Figure \ref{fig:bbpcn} shows the bandwidth blocking probability for constant traffic against various slot sizes. Here the conclusions are same for as the blocking probability, except bandwidth blocking probability is slightly higher for all the slot sizes; because instead of counting the number of blocked connections, we are taking into account the bandwidth of each lightpath request.

In figure \ref{fig:secn} spectrum efficiency is compared for different network topologies for multiple slot widths. The spectrum efficiency is higher when the blocking probability is low due to the minimum wastage of the bandwidth assigned. The slot width 6.25 GHz, 12.5 GHz, 25 GHz, 50 GHz and 100 GHz has higher spectrum efficiency. Even for the single link case at these points, the spectrum efficiency of $100 \%$ is achieved.

\begin{figure}[h]
     \centering
     \captionsetup{justification=centering}

     \begin{subfigure}[b]{0.2\textwidth}
         \centering
         \includegraphics[width=\linewidth]{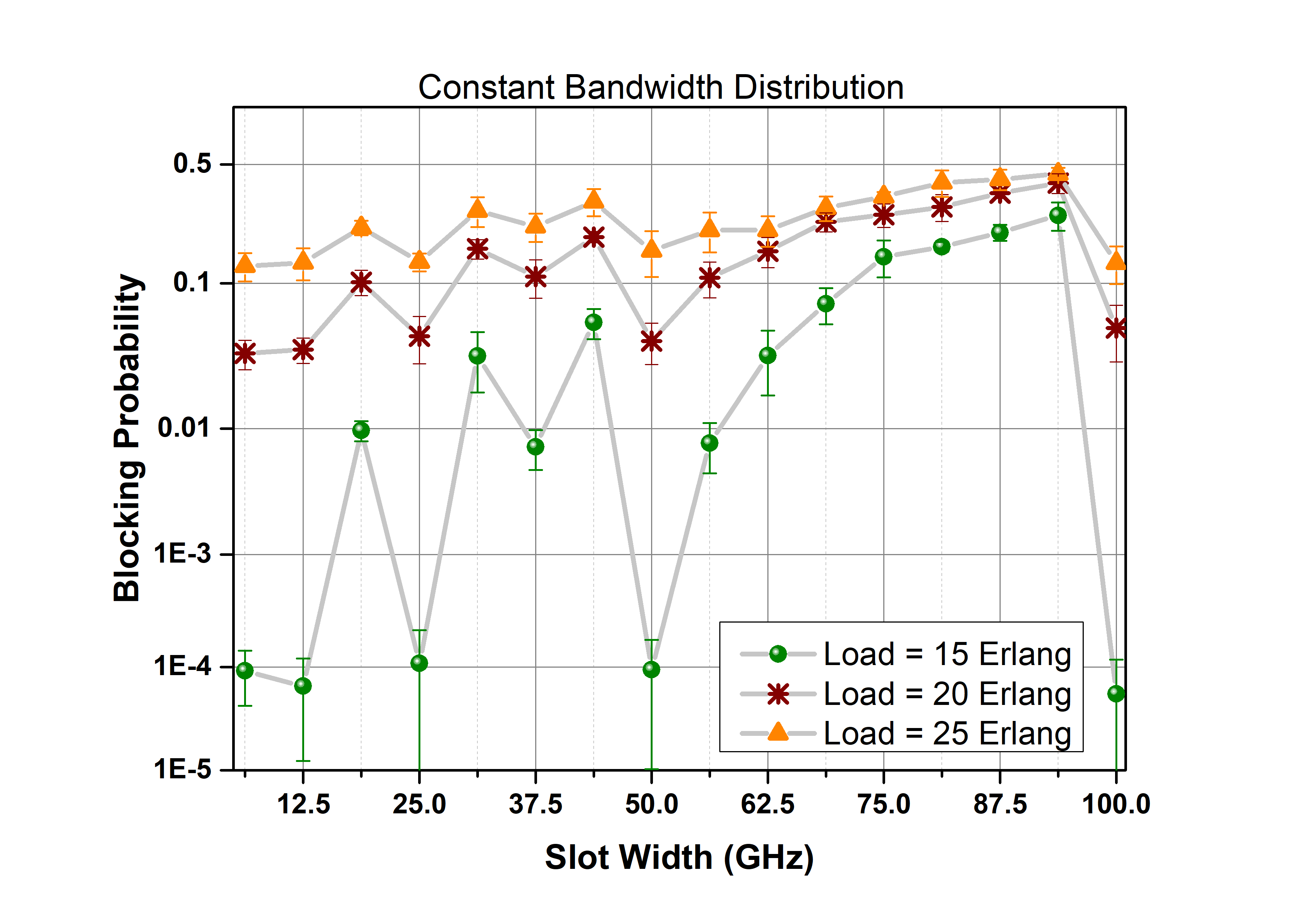}
         \caption{Blocking Probability against various Slot Width.}
         \label{fig:bpcl}
     \end{subfigure}
     \hfill
     \begin{subfigure}[b]{0.2\textwidth}
         \centering
         \includegraphics[width=\linewidth]{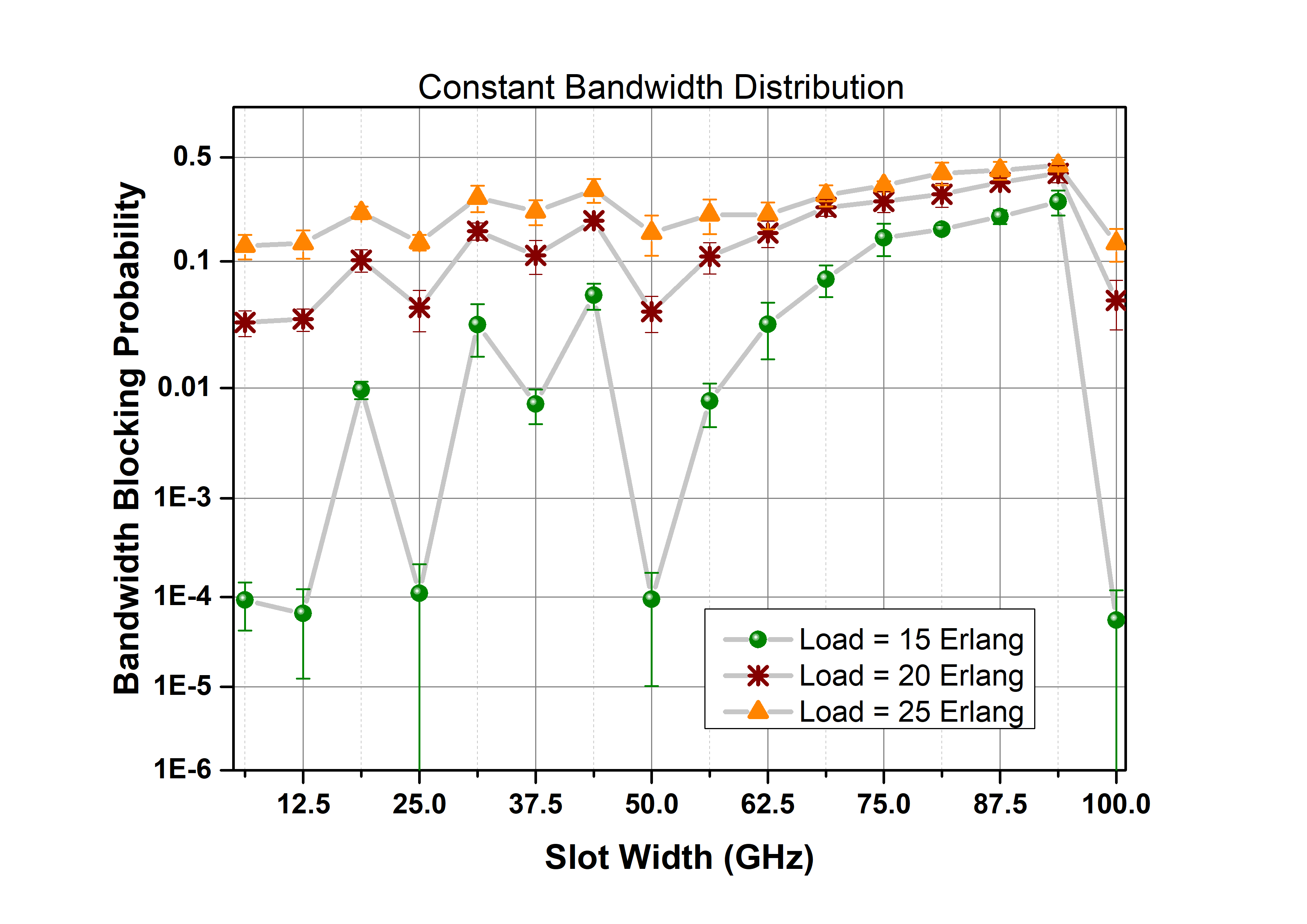}
         \caption{Bandwidth Blocking Probability against various Slot Width.}
         \label{fig:bbpcl}
     \end{subfigure}
     \begin{subfigure}[b]{0.2\textwidth}
         \centering
         \includegraphics[width=\linewidth]{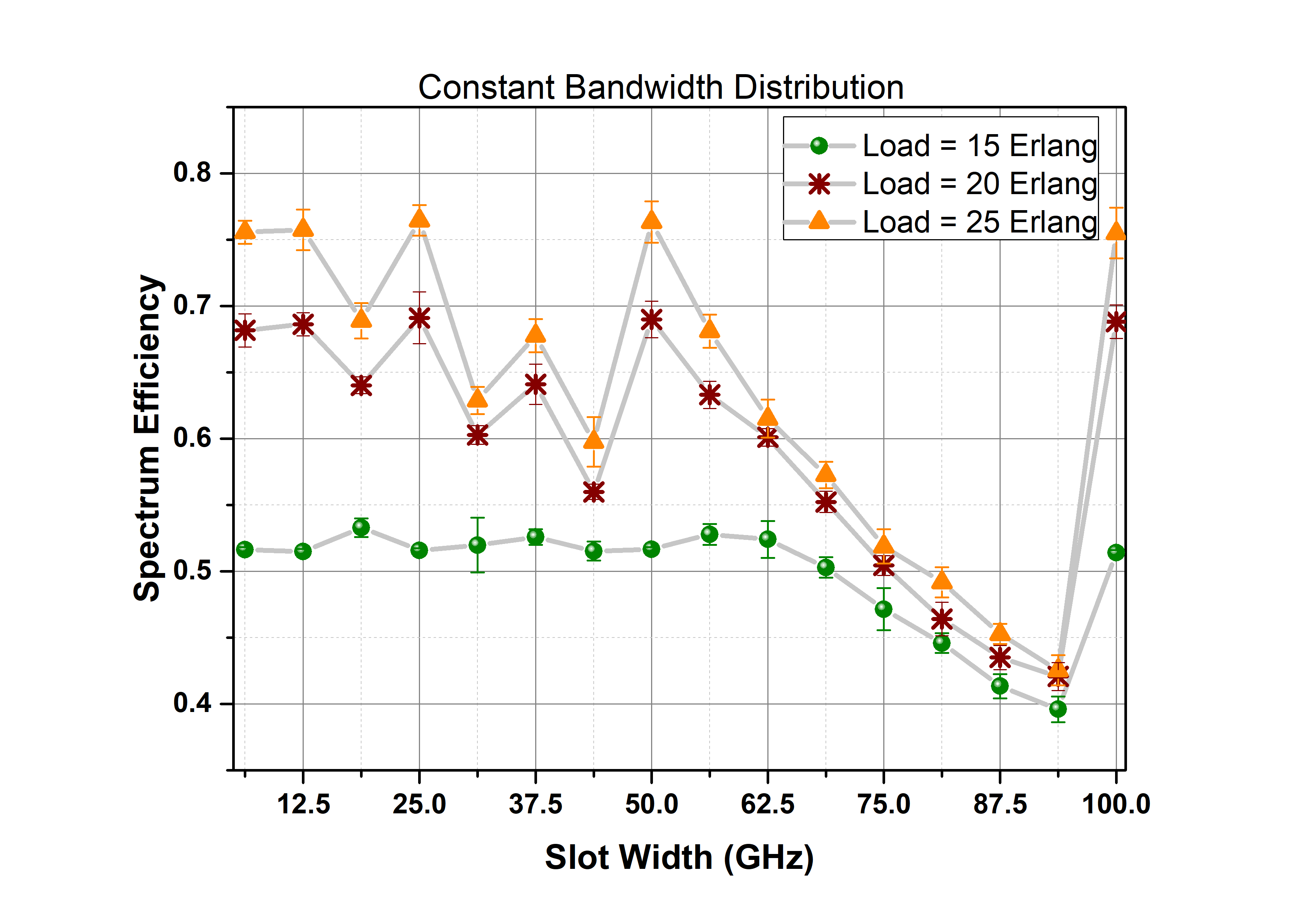}
         \caption{Spectrum Efficiency against various Slot Width.}
         \label{fig:secl}
     \end{subfigure}
        \caption{Various performance parameters for different Offered Load per Node, considering $B$ to be 100 Gbps and NSFNET network topology.}
        \label{fig:consl}
\end{figure}

The same patterns are observed for different loads. Figure \ref{fig:consl} shows plots of different performance parameters against various slot widths. In these plots, we have taken different offered loads per node (in Erlangs). The blocking probability, bandwidth blocking probability, and spectrum efficiency increase as the offered load per node increases. 

In these plots also 6.25 GHz, 12.5 GHz, 25 GHz, 50 GHz and 100 GHz have the lower blocking probability, and higher spectrum efficiency.

Based on the observation, the slot width of 100 GHz outperforms every other options. For constant traffic, if the lightpath request is of size \textit{B} Gbps, the \textit{n} GHz (where $n = B$) slot width outperforms because there is no impact of contiguity constraint. Also, the overhead within the nodes of the network will be low.

\section{Conclusion}

Elastic bandwidth slots can be formed in Flexible Grid Optical Networks using Optical Orthogonal Frequency Division Multiplexing (O-OFDM). It allows dynamic spectrum use by allocating integral multiple of a single slot to the lightpath requests. The slot width according to ITU-T G.694.1 is 12.5 GHz. In this paper, we figured out the most significant slot width for different bandwidth distributions.  We can say that there is an impact of slot width size on the performance of the network. 1 GHz is the optimal slot size due to lower blocking probability, and more spectrum can be accessible for connection accommodation. But, this will increase the overhead within the nodes of the network. The slot size of 6.25 GHz and 12.5 GHz are near-optimal and performs appropriately for the considered bandwidth distributions, slot widths, and offered load per node for Uniform and Poisson bandwidth distribution. Whereas, for Constant bandwidth of \textit{B} Gbps, a slot width of \textit{n} GHz (where $n = B$) outperforms  all other options because there is no impact of contiguity constraint. Hence the fragmentation is low.


\begin{thebibliography}{9}

\bibitem{cisco}
Cisco Annual Internet Report (2018–2023) White Paper, \textit{\url{https://www.cisco.com/c/en/us/solutions/collateral/executive-perspectives/annual-internet-report/white-paper-c11-741490.html}}.

\bibitem{itut}
ITU-T G.694.1, "spectrum grids for WDM applications: DWDM frequency grid", 2012.

\bibitem{eon}
Gerstel \textit{et. al}, "Elastic Optical Networking: A New Dawn for the Optical Layer?", \textit{ in IEEE Communications Magazine}, vol. 50, no. 2, pp. s12-s20, February 2012.
\bibitem{sw1}
Castro \textit{et. al}, "Dynamic routing and spectrum (re)allocation in future flexgrid optical networks", \textit{Computer Networks}, vol. 56, no. 12, pp.2869-2883, 2012.
\bibitem{sw2}
Ujjwal \textit{et. al}, "Optimal spectrum Slot Width Assignment in Flexible Grid Elastic Optical Network",  \textit{2018 3rd International Conference on Microwave and Photonics (ICMAP), Dhanbad},  pp. 1-2, 9-11 February, 2018.

\end{thebibliography}
\end{document}